\documentclass[final,3p,times]{elsarticle}

\usepackage{amssymb}
\usepackage{amsthm}
\usepackage{amsmath}
\DeclareMathOperator*{\argmax}{arg\,max}
\usepackage{units}
\usepackage{verbatim}
\usepackage{adjustbox}
\usepackage{longtable}
\usepackage{multirow}
\usepackage{url}
\usepackage{color}
\usepackage{floatrow}
\usepackage{subcaption}
\graphicspath{ {images/} }

\journal{Applied Energy}

\begin{document}

\begin{frontmatter}

\title{Game-theoretic modeling of curtailment rules and network investments with distributed generation\tnoteref{t1}}

\author[label1]{Merlinda Andoni \corref{cor1}}
\ead{ma146@hw.ac.uk}
\address[label1]{Institute of Sensors, Signals and Systems, Heriot-Watt University, Edinburgh, UK}
\author[label1]{Valentin Robu}
\ead{v.robu@hw.ac.uk}
\author[label2]{Wolf-Gerrit Fr{\"u}h}
\ead{w.g.fruh@hw.ac.uk}
\address[label2]{Mechanical, Process and Energy Engineering, Heriot-Watt University, Edinburgh, UK}
\author[label1]{David Flynn}
\ead{d.flynn@hw.ac.uk}

\cortext[cor1]{School of Engineering and Physical Sciences, Earl Mountbatten Building 3.31, Gait 2, Heriot-Watt University, EH14 4AS Edinburgh, UK}

\begin{abstract}
Renewable energy has achieved high penetration rates in many areas, leading to curtailment, especially if existing network infrastructure is insufficient and energy generated cannot be exported. In this context, Distribution Network Operators (DNOs) face a significant knowledge gap about how to implement curtailment rules that achieve desired operational objectives, but at the same time minimise disruption and economic losses for renewable generators. In this work, we study the properties of several curtailment rules widely used in UK renewable energy projects, and their effect on the viability of renewable generation investment. Moreover, we propose a new curtailment rule which guarantees fair allocation of curtailment amongst all generators with minimal disruption. Another key knowledge gap faced by DNOs is how to incentivise private network upgrades, especially in settings where several generators can use the same line against the payment of a transmission fee. In this work, we provide a solution to this problem by using tools from algorithmic game theory. Specifically, this setting can be modelled as a Stackelberg game between the private transmission line investor and local renewable generators, who are required to pay a transmission fee to access the line. We provide a method for computing the equilibrium of this game, using a model that captures the stochastic nature of renewable energy generation and demand. Finally, we use the practical setting of a grid reinforcement project from the UK and a large dataset of wind speed measurements and demand to validate our model. We show that charging a transmission fee as a proportion of the feed-in tariff price between 15\%-75\% would allow both investors to implement their projects and achieve desirable distribution of the profit. Overall, our results show how using game-theoretic tools can help network operators to bridge the knowledge gap about setting the optimal curtailment rule and determining transmission charges for private network infrastructure.
\end{abstract}


\tnotetext[t1]{This paper builds on significant extensions, both in theoretical results and new datasets, of preliminary work presented at two international conferences: AAMAS 2016~\cite{andoni2016using} and IEEE ISGT Europe 2016~\cite{andoni2016game}.}

\end{frontmatter}

\section{Introduction}
\label{Intro}
Renewable energy is crucial for achieving our decarbonisation goals and mitigating climate change. The Paris Agreement charts a new course of international effort to combat climate change with 195 countries agreeing to keep average global temperature rise well below $2^{\rm{o}}\rm{C}$ above pre-industrial levels and 129 countries ratifying so far. Driven by national and global initiatives, financial incentives and technological advances have permitted large volumes of renewable energy sources (RES) to be connected to the electricity grid. In 2015, 147 GW of new renewable generation capacity was added globally, including 50 GW of new solar PV and 63 GW of wind power capacity, with total investment reaching an estimate of \$285.9 billion \cite{REN21}. The levelised cost of energy (LCOE) for several RES technologies, such as onshore wind or large scale PV, is currently competitive with conventional generation \cite{BEIS2016}. Renewable generation can provide benefits to network operators and consumers, but when installed with high penetration level, it might have negative effects on the operation, resilience and safety of the electricity grid. RES are intermittent and have variable power outputs due to constantly changing primary resources and weather patterns, which are difficult to predict. The challenges faced by network operators relate to reverse power flows, increased power losses, harmonics, voltage fluctuations, thermal capacity of equipment, frequency and voltage regulation and can compromise the system reliability \cite{niemi2010decentralised}.

An additional barrier is that grid infrastructure is inadequate to support continuous RES development or distributed generation (DG), especially in the area of distribution networks. Often high investment takes place in remote areas of the grid, where projects face favourable resource conditions and planning approval. Typically, in the UK, such areas are windy islands or peninsulas with limited or saturated connection to the main grid, facing network constraints, such as voltage, frequency or fault level violations in the absence of a network upgrade. Examples include the Shetland and Orkney archipelagos and the Kintyre peninsula, used as a case study to validate the model developed in this work.

As RES penetration increases, electricity grids require flexible services, which ensure safe operation and power system stability, such as forecasting techniques for RES output prediction, demand response, energy storage and generation curtailment. Most measures can be expensive, such as storage, technically challenging, or not yet mature enough for commercialisation. Hence, in light of the aforementioned barriers, the network places heavy dependence on curtailment at the present time.

Generation curtailment occurs when the excess energy that could have been produced by RES generators is wasted, as it cannot be absorbed by the power system and it cannot be transported elsewhere. In several countries including the UK, generators are compensated for lost revenues, but this results in higher system operation costs which are eventually passed on to end-consumer electricity prices. As more RES continue to be deployed, this practice cannot be sustainable and cost-effective, therefore smart solutions are required for further RES integration.

The two main techniques explored by network operators are Dynamic Line Rating (DLR) and Active Network Management (ANM). DLR uses rating technology and instrumentation to monitor the thermal state of the lines in real time and may improve the estimated capacity between 30\% to 100\% \cite{DOE2012, michiorri2011dynamic}. ANM is the automatic control of the power system by means of control devices and measurements that allow real time operation and optimal power flows. DLR and ANM can be combined to provide greater benefits in terms of curtailment reduction \cite{anaya2014experience}.

From the DNO perspective, both techniques imply controlling DG power outputs, hence innovative commercial agreements between generators and the system operator are required. Generators are offered interruptible, `non-firm' connections to the grid, along with a set of rules about the order they are dispatched or curtailed, as opposed to traditional `firm' connections, a solution preferred in many occasions to avoid high costs or enduring a long wait before getting connected \cite{currie2011commercial}. These terms and conditions are known as `Principles of Access' (PoA) and are the focus of this work. Such schemes have been supported by the UK Government through funding mechanisms encouraging DNOs to facilitate renewable connections \cite{lowcarbon}.

The PoA options chosen by DNOs follow different approaches and each rule has both advantages and disadvantages in achieving desired objectives, such as cost-effectiveness, economic efficiency and social optimality \cite{anaya2014experience}.
A review on different rules is provided in Section~\ref{Com_Agr} and related research works and discussion in Section~\ref{Rel_Work}. DNOs face the knowledge gap of implementing those curtailment rules that achieve greater benefits for all parties involved (RES generators and system operator). However, few works focus on the impact of different rules on the profitability of RES generators, crucially also affecting the investors' decision-making on future generation expansion. Our work studies the effects of different rules on the viability of RES developments. In particular, we provide results based on simulation analysis that show how several rules can decrease the capacity factor (CF) of different wind generators and how correlated wind speed resources affect the resulting curtailment.

The main rules found in the literature or applied in practical settings follow either a Last-in-first-out (LIFO) or a Pro Rata approach. LIFO is easily implemented and does not affect existing generators, but might discourage further RES investment. On the other hand, Pro Rata shares curtailment and revenue losses equally amongst all generators, who face frequent disruption every time curtailment is required. Note here that the fairness property plays a key role in maximising the renewable generation capacity built \cite{andoni2016using} and can lead to higher network utilisation \cite{BAR12}. Inspired by simultaneously satisfying objectives such as fairness, not discouraging future RES development and minimising disruption, we propose a new `fair' rule which reduces the curtailment events per generator and guarantees approximately equal share of curtailment to generators of unequal rated capacity.

While smart solutions can defer network investment, the implications of curtailment extend to inefficient energy management and renewable utilisation, potential economic losses for RES generators, wasted energy and increased operation and balancing costs. Future adoption of electric vehicles (EVs)~\cite{druitt2012simulation, robu2011online, stein2012model} and battery energy storage systems could be used to store excess RES generation and reduce curtailment. A long term sustainable solution to facilitate low-carbon technologies is  network upgrade, such as reinforcing or building new transmission/distribution infrastructure. Grid expansion is a capital intensive investment, traditionally performed by system operators and heavily subsidised or supported by public funds. According to \cite{bronski2015}, the USA grid capacity investment would require an estimated \$100 billion per year, between 2010 and 2030, with a minimum of \$60 billion related only to the integration of RES. In the UK, an estimated £34 billion of investment in electricity networks could be required from 2014-2020, to accommodate new onshore and offshore projects \cite{DECC2015}. Deregulated electricity markets and RES integration enable private investors participation in network investments. This market behaviour can be desirable from a public policy standpoint but it raises the question for system operators of defining the framework within which these private lines are incentivised, built and accessed by competing generators. Currently, DG investors bear a part of the costs required for their integration. In general the connection costs may vary, but usually include the full cost for the grid capacity installed for own use and a proportion of the costs for shared capacity with other customers, in the case of a network upgrade \cite{anaya2015options}. The remaining costs are recovered by the system charges borne by all grid users, representing approximately 18\% of the average electricity bill of a typical UK household \cite{ward2012demand}.

Moreover, current practices may lead to inefficient solutions in real-world settings, such as the problem of reinforcing transmission/distribution lines in outlying regions of Scotland, such as in the Kintyre peninsula, an area that has attracted major RES investment and is used as a case study in this paper. The scheme providing grid access to the RES generators in this area followed a `single access' principle, i.e. private lines for sole-use that were sufficient to accommodate only the RES capacity of each project. This practice resulted in the unintended effect of no less than three lines being connected or under construction in the same area. It is clear that current solutions are far from being optimal in terms of network use and economic efficiency.

In similar settings, it is possible for system operators to encourage RES generators to install larger capacity power lines under a `common access' principle, where a private investor is granted a license to build a line that permits access to smaller generators, who are subject to transmission charges. In these settings, curtailment and line access rules can play a significant role in the resulting strategic expansion. We use tools from algorithmic game theory to model this complex interplay and help DNOs or the system regulator to optimally determine the transmission charges that enable private infrastructure be installed and help the process of building efficient and resilient networks. In the recent years, game-theoretic tools have gained increased popularity within the Energy and Power System community. We provide relevant works using hierarchical games in the context of network upgrade settings in more detail in Section~\ref{Rel_Work}. While other works have studied transmission constraints and congestion, to the best of our knowledge this work is one of the first to study the effect of commercial agreements and curtailment rules in settings of private grid reinforcement. Our work provides a novel formulation in modeling private investment in power network infrastructure required to further integrate renewable generation. We show how different curtailment rules affect investors' decision making about additional generation and transmission capacity, providing a useful tool to network operators seeking to incentivise sustainable and low-carbon technologies. In more detail, the main contributions of our work to the existing state-of-the-art are:
\vspace{-2mm}
\begin{itemize}
\item We provide a study for three curtailment rules and show simulation analysis results about their impact on the capacity factor of wind generators. We also study the effect of \textit{spatial wind speed correlation} to the resulting curtailment and lost revenues. The results provide useful insights to DNOs searching to implement DG smart connections and optimal curtailment rules. A new curtailment strategy is proposed, which is fair and causes minimal disruption.
\vspace{-2mm}
\item This work develops a game-theoretical model which enables network operators to bridge the knowledge gap of incentivising privately developed grid infrastructure, especially in settings where multiple generators can share access through the same transmission line, and determining suitable transmission charges. The network upgrade, under a `common access' principle and a fair curtailment rule, is modelled as a \textit{Stackelberg game} \cite{von2010market} between the line investor and local generators. Stackelberg equilibria are classified as solutions to sequential hierarchical problems where a dominant player (here the line investor) has the market power to impose their strategies to smaller players (local generators) and influence the price equilibrium. The equilibrium of the emerging game determines the optimal generation capacity installed and resulting profits for both players under varying cost parameters. A feasible range of the transmission fee is identified allowing both transmission and generation capacity investments be profitable.
\vspace{-2mm}
\item Finally, the theoretical analysis is applied to a real network upgrade problem in the UK. The datasets used include real wind speed measurements and demand data that span over the course of 17 years. This case study analysis demonstrates how real data can be utilised to search and identify the equilibrium of the game with an empirical approach, that allows capturing the \textit{stochastic} nature of wind generation and varying demand.
\end{itemize}
\vspace{-2mm}

The remainder of the paper is organised as follows: Section~\ref{Rel_Work} reviews relevant literature, Section~\ref{Com_Agr} presents most important PoAs and proposes a new curtailment rule, Section~\ref{Net_up} presents the network upgrade model, Section~\ref{Cas_Stud} analyses a practical case study, Section~\ref{Res} discusses our main findings and Section~\ref{Con_Fut} concludes this work.

\section{Related work}
\label{Rel_Work}
The literature review presented refers to related work about curtailment rules and smart solutions, network expansion and the use of game theory in the energy field.

\subsection{Literature review on curtailment rules}

The rules for curtailment are specified in the legally binding agreement between the RES generator and the system operator. An extensive review of different rules can be found in \cite{anaya2015options, kane2015evaluation}. A number of commercial and academic studies \cite{currie2011commercial, BAR12, kane2015evaluation, bell2011academic, eirgrid2011ensuring} have discussed issues around the application of these strategies, with main focus on their technical, legal and regulatory implications. However, few research works have focused on their effects on the profitability of RES generators. Along with other financial incentives provided to renewables, such as the level of feed-in-tariff prices, the curtailment rule selected in PoAs and the curtailment level are key factors affecting the investors decision-making on future projects. Our work focuses on the impact of different rules on the viability of RES investment and the decision-making of investors about future generation expansion.

Anaya \& Pollitt \cite{anaya2015options} provided a cost-benefit analysis which compares traditional connections with network upgrade to smart interruptible connections. Their results are based on static assumptions of the generation mix and curtailment levels. The results from our work are based on hourly RES resources and demand.

The main threads found in the literature are Last-in-first-out rules which do not affect existing generation, Pro Rata rules that share curtailment equally amongst all generators, or Market-based rules that require the establishment of a curtailment market. These rules were discussed in \cite{anaya2014experience} in terms of risk allocation and social optimality, rather than their effect on the viability of RES investments, which is the focus of our work.

Similar to \cite{kane2015evaluation}, our work takes a direct approach in quantifying the effects of most commonly used PoAs to the capacity factor of wind generators by a simulation analysis. In addition, our work demonstrates how wind speed spatial correlation affects the resulting curtailment and how different PoAs affect the frequency of curtailment events, providing useful insights to DNOs regarding the most efficient strategy. Correlation should not be ignored as most generators responsible for a particular grid constraint have geographical proximity and therefore correlated power outputs, resulting in a greater impact of the resulting curtailment.

Finally, this work extends previous work by providing a model that captures the stochastic nature of renewable resources and variation in demand, as opposed to the average analysis approach presented in \cite{andoni2016using, andoni2016game}.

\subsection{Review of game theory and artificial intelligence techniques applied to Energy Systems}
In the context of deregulated electricity market, several authors have argued that transmission planning techniques need to adopt optimisation \cite{ruiz2014tutorial} and strategic modelling of market participants \cite{day2002oligopolistic}, as opposed to `rules of thumb' approaches driven by human management experience, traditionally performed by utility companies in the past \cite{hobbs1995optimization}. These techniques include mathematical optimisation techniques or game theory models.

Significant factors to take into account include uncertainties introduced by distributed resources and renewable generation, requiring increased network upgrade investment and decreasing network assets utilisation. Many works focus on planning expansion techniques incorporating multi-objective optimisation, such as in \cite{foroud2010multi, soroudi2010distribution} focusing on distribution expansion and in \cite{arabali2014multi} where the optimisation criteria considered were the investment and congestion costs, and the system's reliability. Akbari et al. \cite{akbari2011multi} provided a stochastic short term transmission planning model based in Monte Carlo simulations, while Zeng et al. \cite{zeng2016multi} considered a multi-level optimisation approach for active distribution system planning with renewable energy harvesting, taking into consideration reinforcement and operational constraints. In \cite{fini2013investigation, kamalinia2014sustainable}, the authors studied distributed generation expansion planning with game theory and probabilistic modelling with strategic interactions, respectively. Other works consider an integrated model for both generation and transmission capacity \cite{baringo2012transmission, gupta2014computationally} or in \cite{motamedi2010transmission} where the effect of generation capacity on transmission planning was examined.

Multi-objective optimisation techniques were utilised for the integration of renewable energy sources in order to achieve optimal design of renewable systems at a microgrid level \cite{baghaee2016reliability} or stand-alone systems using particle swarm optimisation \cite{kaviani2009optimal} with focus on the system's reliability \cite{baghaee2016multi}.

Several works have discussed private investment in the field of grid infrastructure. Contreras et al. \cite{contreras2009incentive} introduced an incentive scheme based on the Shapley value to encourage private transmission investment. Maurovich-Horvat et al. \cite{maurovich2015transmission} compared two alternate market structures for grid upgrades (either by system operators or private investors) and showed that they can lead to different optimal results. However, their work was focused on using Mathematical Programming with Equilibrium Constraints (MPEC) to solve power flows and curtailment strategies were not considered. Perrault \& Boutilier \cite{perrault2014efficient} used coalition formation to coordinate privately developed grid infrastructure investments with the aim to reduce inefficiencies and transmission losses. The main focus of their work was the group formation and its effects on configurations of multiple-location settings, not the effects of line access rules and smart solutions which forms the scope of this work. Joskow and Tirole \cite{joskow2000transmission} analysed a two-node network market behaviour, for settings of players with different market power and allocation of transmission rights at congested areas of the power network. Our work follows a different approach, since we specify our analysis on the transmission access rules and curtailment imposed. Grid expansion in a national level was studied in \cite{huppmann2015national} as a three-stage hierarchical Nash game.

Stackelberg games have been used in several works for transmission upgrade. These works considered economic analysis with social welfare \cite{sauma2006proactive}, Locational Marginal Pricing \cite{shrestha2004congestion} or highlight the uncertainties of RES generation \cite{van2012economics}. Recent work in the renewable energy domain used Stackelberg game analysis to study energy trading among microgrids \cite{asimakopoulou2013leader, lee2015distributed}.

In recent years several researchers have begun to show the benefits of game-theoretic, multi-agent modelling and artificial intelligence (AI) techniques applied to power markets, including for integration of distributed, intermittent renewable generations resources. One such prominent example in the multi-agent and AI community was PowerTac \cite{ketter2013power}. Baghaee et al.~\cite{baghaee2016three, baghaee2017nonlinear, baghaee2017application} used artificial neural networks to model probabilistic power flows in microgrids with increased RES penetration. Game-theoretical analysis was utilised in smart grid settings with demand-side management \cite{kota2012cooperatives, lidemand, ma2016incentivizing, xudemand}, Virtual Power Plants \cite{robu2012cooperative, robu2016rewarding, vasirani2013agent} and cost-sharing of generation and transmission capacity \cite{perrault2014efficient}. Both non-cooperative \cite{su2014game} and cooperative \cite{zhang2015game} game theoretical models were used to clear deregulated electricity markets or were used to model the operation of microgrids \cite{prete2016cooperative}. Min et al.~\cite{min2013game} defined the generators' strategies by Nash equilibria. Wu et al. \cite{wu2016profit} discussed coalition formation and profit allocation of RES generators within a distributed energy network comprised of controllable demand. Zheng et al. \cite{zhengcrowdfunding} proposed a novel, crowdsourced funding model for renewable energy investments, using a sequential game-theoretic approach. However, the issue of investing in transmission/distribution assets was not considered. Game theory techniques for distribution network tariffs determination were discussed in \cite{ortega2008distribution} with the objective of maximising social welfare.

Finally, this paper extends our previous work initiated in \cite{andoni2016using, andoni2016game}, in such way that stochastic renewable generation and varying demand are captured in the equilibrium results.

\section{Fractional Round Robin as a new curtailment rule}
\label{Com_Agr}

In this section we elaborate on the most widely used curtailment rules found in the literature or applied in commercial schemes. We also propose a new curtailment rule and demonstrate its advantages through a simulation process over the course of one year.
 
\subsection{Principles of Access}
\begin{table}
\centering
\begin{adjustbox}{width=1                                                    \textwidth,center=\textwidth}
\small
\begin{tabular} { lll}
 \hline
 \hline
 \textbf{~Principle of Access/Curtailment order} & \textbf{~Advantages}  & \textbf{~Disadvantages} \\
 \hline
 \hline
 \multirow {4}{10em}{LIFO: Last generator} & Simple configuration & Not equitable \\
 & No impact on existing connections & Not favourable to RES generators\\
 & Easily computable capacity factor & Inefficient use of distribution network\\
 & Consistent and transparent & Generation capacity disincentivisation\\
 \hline
 Rota: Rotationally & Smaller capacity factor reduction with increased units & Greater impact for small-sized generators\\
\hline
 \multirow{3}{10em}{Pro Rata: According to rated capacity or power output} & Equitable & Unknown long term impacts on capacity factor\\
 & Compliant with existing rules and standards & Increased curtailment for additional units \\
 & Enhances competitiveness & Possible need cap generation connected\\
\hline
\multirow{3}{10em}{Market Based: Bidding for grid access or curtailment} & Sustainable and future proof & Necessity of market development\\
& No impact on existing connections & More suitable for transmission than distribution networks\\
& Enhances competitiveness & \\
\hline
\multirow{3}{10em}{Greatest Carbon Benefit: Larger CO2 emissions generator} & Easy technical implementation & Commercial implications\\
& & Not easy to determine carbon footprint\\
& & Needs regulatory changes\\
\hline
Generator Size: Larger generator & Quick removal of constraint & Possible discouragement of large generators\\
\hline
Technical Best: Most technically suitable & DNO Encouragement for grid reinforcement & Location dependable\\
\hline
\multirow{2}{10em}{Most convenient: Most likely to respond} & Simple method & Discrimination on operator preference, size and location\\
& & Not fair, not transparent\\
\hline
\hline
\end{tabular}
\end{adjustbox}
\caption{Main features of common curtailment strategies: \cite{currie2011commercial} provided an assessment of different PoAs for interruptible contracts with respect to different criteria and stakeholders, \cite{bell2011academic} identified a range of different criteria for assessing the most suitable PoA for ANMs, \cite{eirgrid2011ensuring} focused on technical challenges caused by increased wind penetration and \cite{kane2014review} provided a comprehensive review on PoAs quality assessment for ANM settings}
\label{TabPoA}
\end{table} 

While a larger number of curtailment rules is summarised and reviewed in Table~\ref{TabPoA}, in this work we focus our attention on  three main schemes, which are mainly applied in commercial projects in the UK and other countries:
\vspace{-2mm}
\begin{itemize}
\item \textit{LIFO:} Last-In-First-Out rule curtails first the generator that was connected last in the ANM scheme. Early connections have a clear market advantage. The LIFO rule was selected by Scottish and Southern Energy (SSE) in two occasions as being transparent, simple to implement and not affecting existing generators. The first practical example was the \textit{`Orkney Smart Grid'} scheme\footnote{\url{https://www.ssepd.co.uk/OrkneySmartGrid/}} with a capability of controlling power flows within a time period of a few seconds. In this occasion, each generator is connected to a local ANM controller, which receives a power output set point from the central controller. The ANM scheme cost $\pounds 500\rm{k}$ to install, significantly lower than the $\pounds 30\rm{m}$ estimated cost of upgrading the grid infrastructure, and allowed $20~\text{MW}$ additional capacity. The second example was the \textit{`Northern Isles New Energy Solution (NINES)'} project\footnote{\url{https://www.ssepd.co.uk/NINES/}} which combined smart grid technologies, storage heating and demand side management.
\vspace{-2mm}
\item \textit{Rota:} Generators are curtailed on a rotational basis or at a predetermined rota specified by the system operator. An early example of the Rota rule applied in a practical setting in the United States by Xcel Energy can be found in \cite{fink2009wind}.
\vspace{-2mm}
\item \textit{Pro Rata or Shared Percentage:} Curtailment is shared equally between all non-firm generators proportionally to the rated capacity or actual power output of the generators. Pro Rata was chosen by UK Power Networks for their 3-year \textit{`Flexible Plug and Play'} pilot project\footnote{\url{http://innovation.ukpowernetworks.co.uk/innovation/en/Projects/tier-2-projects/Flexible-Plug-and-Play-(FPP)}} which offered feasible connections to 50~MW of distributed generation, as it  permits larger volumes of generation being connected and enhances network utilisation \cite{UKPN14}.
\vspace{-2mm}
\end{itemize}

The curtailment rules have various effects on generators, system operators and consumers. As imposed curtailment reduces the energy produced by generators, it causes a capacity factor (CF) below the possible CF given the resource alone and lost revenues. Financial implications have the potential to discourage, in the long term, the generation capacity investment at the location where ANM is applied which leads to inefficient use of network resources.

The LIFO scheme discriminates according to the order of connection, which might disincentivise future renewable development and makes inefficient use of the transmission capacity available. The Rota scheme is a simple approach which does not take into account the size of the generator or its actual contribution to the network constraint. This results in disproportionate losses of revenue, especially to smaller sized generators. Finally, Pro Rata shares curtailment equally and is fair, however all participating generators are curtailed at all times when curtailment is required, leading to increased disruption. Pro Rata might not always be desirable (technically speaking, it may require modified pitch-controlled wind turbines, such that their output can be adjusted as needed, which may be more expensive), as in several occasions, it is technically preferable to curtail a larger amount of power from one generator than smaller amounts from all generators at a single event. Moreover, `fairness' is significant as fair schemes maximise the generation capacity built at a single location \cite{andoni2016using}.  

For the above reasons, we propose an equivalent Rota-type strategy with `fairness' properties, a strategy called \textit{Fractional Round Robin} (FRR). With FRR, the power curtailed is distributed sequentially on a rotation basis, according to the number of rated capacity units installed, so that larger generators are chosen proportionally more times, in direct relation to their size. The benefits of this approach are better explained with an illustrative example presented in the next section.

\subsection{Illustrative example network}
\label{Illustration}
To illustrate the effects and operation of the most important curtailment schemes, we consider a simplified example network of three wind generators of $\textstyle {P_{N_1}=7~{\rm MW}}$,~$\textstyle {P_{N_2}=2~{\rm MW}}$ and $\textstyle {P_{N_3}=3~{\rm MW}}$ rated capacity, where the subscript denotes the chronological order of their connection to the power grid. For simplicity, we assume there is no export capability and the demand is constant and equal to $\textstyle {P_{D,t}=6~{\rm MW}, \forall t}$. For a given time interval $\textstyle {t}$, if all generators are producing their nominal output power, a total of $\textstyle {P_{C,t} = 6~{\rm MW}}$ needs to be curtailed. The allocation of curtailment to the generators depends on the scheme selected:
\vspace{-2mm}
\begin{itemize}
\item With LIFO, the third and second generator are completely curtailed and the first is curtailed by $\textstyle {1~{\rm MW}}$.
\vspace{-2mm}
\item When Rota is implemented, the generators take turns, resulting here in $\textstyle {6~{\rm MW}}$ curtailed by the first generator, while others are not affected. In the next curtailment event, the second generator is curtailed, but as required curtailment is not sufficient, the third generator is also completely curtailed and  $\textstyle {1~{\rm MW}}$ is curtailed by the first generator. In the next event, the second generator is first to be curtailed and so on.
\vspace{-2mm}
\item By contrast, with Pro Rata the allowed export is allocated proportionally to the generator's output, resulting in $\textstyle {3.5~{\rm MW}}$, $\textstyle {1~{\rm MW}}$ and $\textstyle {1.5~{\rm MW}}$ curtailed power, respectively.
\vspace{-2mm}
\item With FRR, $\textstyle {6~{\rm MW}}$ will be curtailed by the first generator. The generator will still be the first in the order of curtailment in future events, up until a quota equal to its rated capacity is reached. Therefore, the next time curtailment is required, $\textstyle {1~{\rm MW}}$ will be curtailed by the first generator and the remaining $\textstyle {5~{\rm MW}}$ will be curtailed by the second and third generator. This means that on average, every $\textstyle {12}$ times a curtailment of $\textstyle {6~{\rm MW}}$ is needed, the first generator will be curtailed $\textstyle {7}$ times, the second $\textstyle {2}$ times and the third $\textstyle {3}$ times.
\vspace{-2mm}
\end{itemize}

To further elaborate on the effects of the rules we perform a simulation process showing the impact on the CF of the wind generators.

\subsection{Simulation process}

We implement a simulation process in the course of one year, to compute the capacity factors of the wind generators in the network considered in Section~\ref{Illustration}, but under different schemes. Since network constraints are usually applicable to a particular geographical area of the grid where wind conditions may be similar, the power output of the generators presents a level of spatial correlation, which is significant for the required curtailment level at this area. To model correlation, we apply the technique developed by Fr{\"u}h~\cite{fruh2015local}.  First of all, we generate $\textstyle {8760}$ data points of wind speed $\textstyle {u_{rand,i}}$ for $\textstyle {i=1...3}$ generators, from three \textit{random} and independent samples of a Weibull distribution (one for each generator), using typical UK values of $\textstyle {c=9~{\rm m/s}}$ and $\textstyle {k=1.8}$. Weibull distributions are often used to represent wind speed distributions~\cite{tuller84characteristics, edwards2001level}. We set the wind speed at the first generator's location as a reference $\textstyle {u_{Ref}}$ and we produce random, yet cross-correlated wind data series $\textstyle {u_{i}}$ at each generator's location, by the following equations:
\begin{equation}
u_{i}(t)=c_r\cdot u_{Ref}(t)+(1-c_r)\cdot u_{rand,i}(t)
\end{equation}
\begin{equation}
c_r=\frac{1}{\pi}\cdot \arccos(1-2r)
\end{equation}
where $\textstyle {r}$ is the Pearson's correlation coefficient. The data series are then converted to power outputs, using a generic model of a wind turbine. If the aggregate power at time $\textstyle {t}$ exceeds the power demanded, then curtailment is required, which is  allocated to the generators according to the strategy imposed.

\begin{figure*}
\centering
\begin{tabular}{ccc}
\begin{subfigure}{0.33\textwidth}
\includegraphics[width=\linewidth]{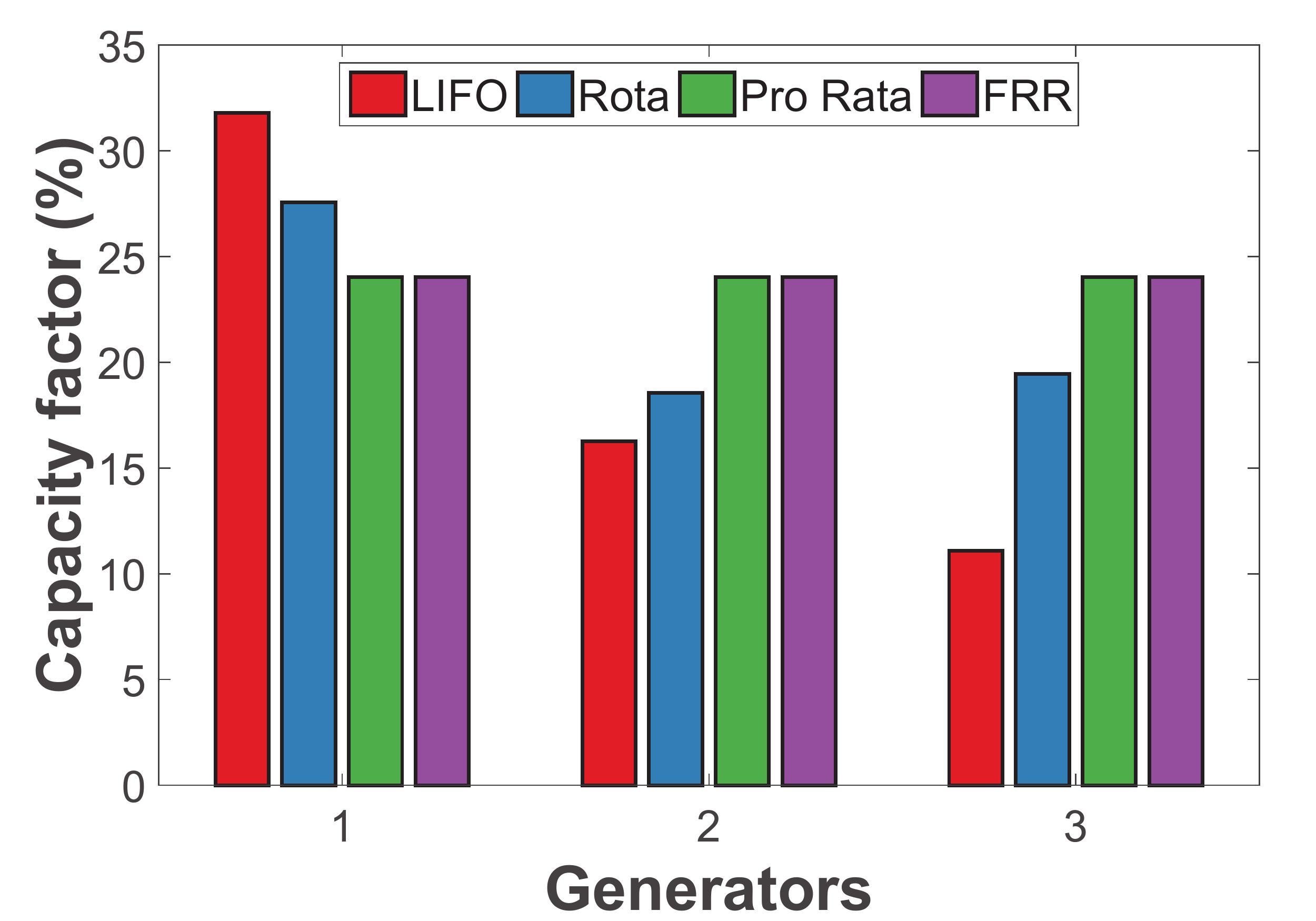} \caption{}
\end{subfigure}
 & 
 \begin{subfigure}{0.33\textwidth}
 \includegraphics[width=\linewidth]{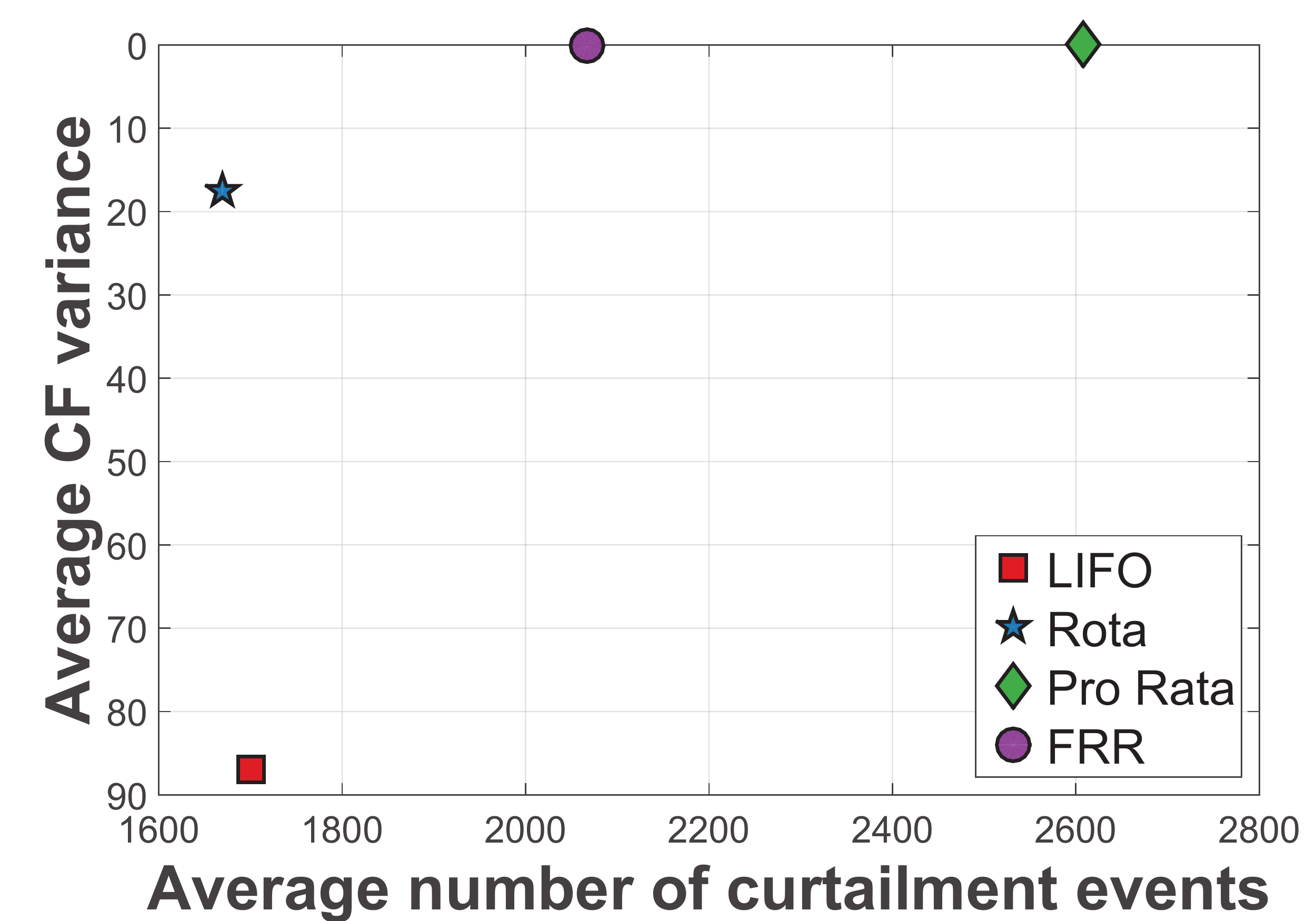} \caption{}
 \end{subfigure} &
 \begin{subfigure}{0.33\textwidth}
 \includegraphics[width=\linewidth]{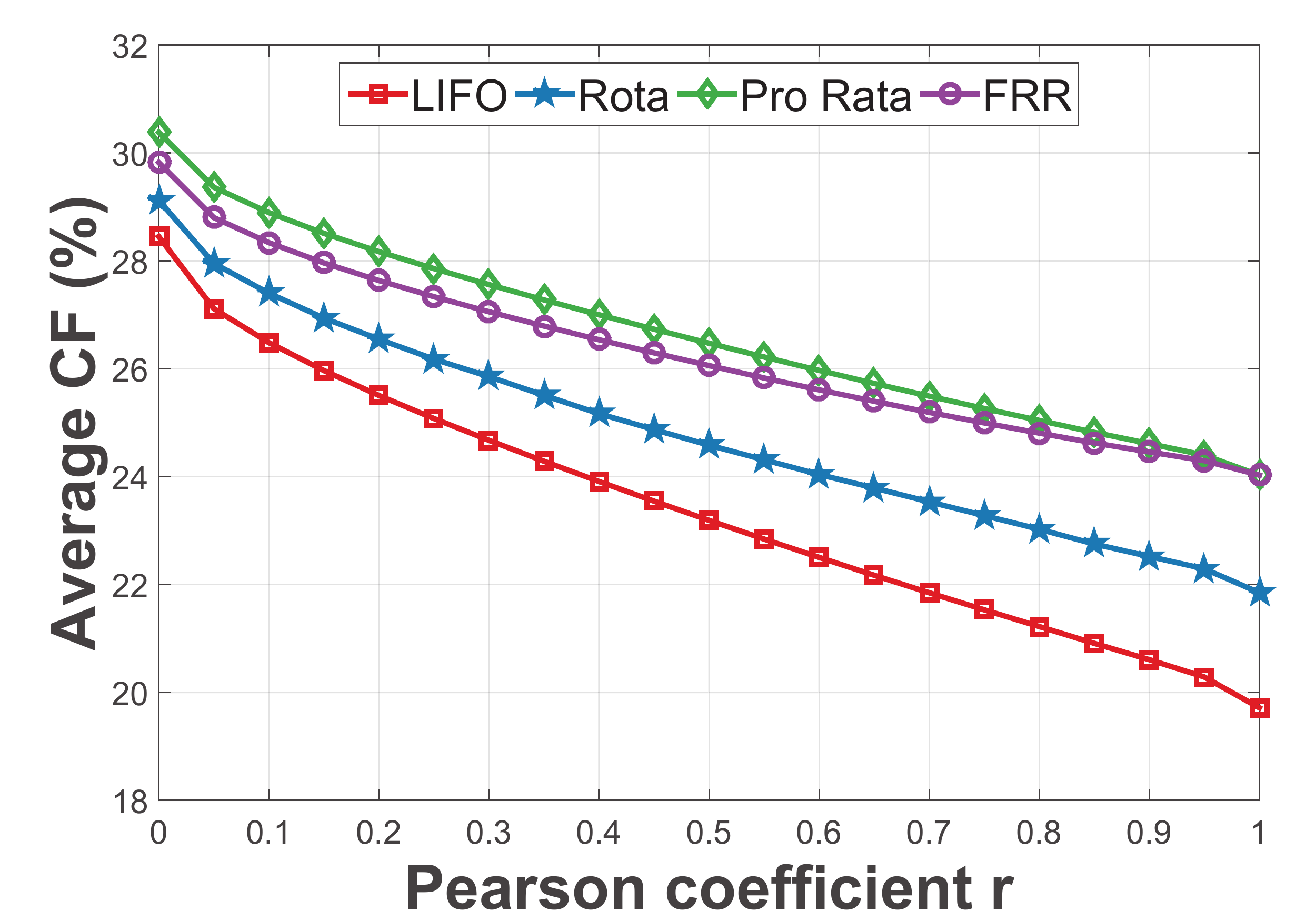} \caption{}
 \end{subfigure}
\end{tabular}
\caption{Curtailment effect on the (a) CF of wind generators, (b) fairness represented by the average CF variance and (c) correlation effects on average CF under LIFO, Rota, Pro Rata and FRR}
\label{CF_Fair_AveCF}
\end{figure*}

Fig.~\ref{CF_Fair_AveCF}a shows the CF results for each generator under the four different schemes for conditions of perfect correlation ($\textstyle {r=1}$). LIFO clearly favours `early' connections, while the third generator suffers a reduction of 67.4\%. As shown, Rota can disadvantage smaller-sized generators. On the contrary, Pro Rata produces equal CF reduction for all generators, while FRR produces similar results to Pro Rata, as expected. A measure of the fairness of a particular strategy is the variance of the capacity factors for the participating stations, $\sigma^2(CF_i)$ where $CF_i$ is the capacity factor for generator $i$. We illustrate this variance with the average number of curtailment events required per generator for the example of $r=1$ in Fig.~\ref{CF_Fair_AveCF}b. LIFO presents a poor performance with respect to fairness, as opposed to Pro Rata, which requires the largest number of curtailment events. Rota is fairer than LIFO, but still presents unequal treatment of generators, and requires the smallest number of curtailment events compared to all schemes. FRR can present similar fairness properties to Pro Rata, while reducing significantly the number of curtailment events per generator. Finally,  as shown in Fig.\ref{CF_Fair_AveCF}c, the required total curtailment increases, as we proceed from no correlation to perfect correlation among the generators, resulting in lower CFs.

Summarising, FRR is a fair strategy which minimises disruption. Moreover, knowledge of the curtailment order in advance reduces the uncertainty of short term power output prediction of a generator. Finally, for a sufficiently long period of time (i.e. many years or the typical lifetime of a wind turbine), the curtailment rate under FRR converges to the proportional curtailment rate with Pro Rata, as shown in Fig.~\ref{CF_Fair_AveCF}a. The remainder of the paper focuses on the network upgrade problem as a solution to reducing curtailment.

\section{Modelling network investment using game theory}
\label{Net_up}
In this section, we examine the combined effects that fair curtailment rules, such as Pro Rata or FRR, and `common access' line rules have on network upgrade and the renewable capacity installed. First, we describe the general setting of the game. Second, we present a model that captures stochasticity of generation and variation in demand.

\subsection{Stackelberg game description}

Consider two locations: $\textstyle {A}$ is a net consumer (where demand exceeds supply, e.g.~a mainland location with industry or significant population density) and $\textstyle {B}$ is a net energy producer (area of favourable renewable conditions, e.g.~a remote region rich in wind resource). In practice, there would be some local demand and supply, considered here negligible, and installation of new RES capacity is not be feasible without a network upgrade.

Moreover, we consider two players: the line investor, who can be merchant-type or a utility company and is building the $\textstyle {A-B}$ interconnection and possibly renewable generation capacity at $\textstyle {B}$, equal to $\textstyle {P_{N_1}}$, and a local player, who represents the local RES generators or investors located at $\textstyle {B}$, $\textstyle {P_{N_2}}$. This second player can be thought of as investors from the local community, who do not have the technical/financial capacity to build a line, but may have access to cheaper land, find it easier to get community permission to build turbines etc., hence may have a lower per-unit generation cost. Note that in Scotland or other countries such as Denmark, local groups often act together to make land available and invest in RES projects. In Scotland, Community Energy Scotland (CES) is an umbrella organisation of such groups and DNOs have an incentive to work with them. $\textstyle {E_{G_i}}$ represents the expected energy units which could be produced over the project lifetime according to the resource on the site's location without encountering curtailment, while $\textstyle {E_{C_i}}$ is the amount of available energy lost through curtailment under the adopted Principle of Access. For simplicity, we assume there is no RES capacity installed at location $\textstyle {B}$ prior to the construction of the power line.

The decision of building the power line will elicit a reaction from local investors. Crucially, the line investor has a \emph{first mover} advantage in building the grid infrastructure, which is expensive, technically challenging, and only a limited set of investors (e.g. DNO-approved) have the expertise and regulatory approval to carry it out. The power line cost is estimated as $\textstyle {C_T=I_T+M_T}$ over the project lifetime, where $\textstyle {I_T}$ is the cost of building the line (or initial investment) and $\textstyle {M_T}$ the cost of operation and maintenance. The monetary value of the power line is proportional to the energy flowing from $\textstyle {B}$ to $\textstyle {A}$, charged for local generators and  common access rules with $\textstyle {p_T}$ transmission fee per energy unit transported through the line. Moreover, for a generation unit $i$, we define the cost of expected generation per unit as $\textstyle {c_{G_i}=(I_{G,i}+M_{G_i})/{E_{G_i}}}$, where $\textstyle {I_{G_i}}$ is the cost of building the plant and $\textstyle {M_{G_i}}$ the operation and maintenance costs, and we assume that the energy generated by a RES unit is sold at a constant generation tariff price, equal to $\textstyle {p_G}$.

Given these parameters, the profit functions of both players are defined. The line investor has two streams of revenue, one from the energy produced and one by the energy produced by other investors or local generators, transported through the transmission line. The costs of the line investor are related to the installation and operation of the generation capacity (generation cost $c_{G_1}$) and the installation of the power line $C_T$.
\begin{equation}
\Pi_1=(E_{G_1}-E_{C_1})p_G-E_{G_1}c_{G_1}+(E_{G_2}-E_{C_2})p_T-C_T
\label{Prof_1}
\end{equation}
Similarly, the profits of the local generators $c_{G_2}$ depend only on the energy they produce with some generation cost $c_{G_2}$ and then transmit through the power line at an access charge of $p_T$:
\begin{equation}
\Pi_2=(E_{G_2}-E_{C_2})(p_G-p_T)-E_{G_2}c_{G_2}
\label{Prof_2}
\end{equation}

The research question is \textit{`How to determine the optimal generation capacities built by the two players, so that profits are maximised?'} If we view this from a game-theoretic perspective, this is a bi-level Stackelberg game \cite{von2010market}. The line investor or \emph{leader} can assess and evaluate the reaction of other investors to determine their strategy (i.e. the generation capacity to be installed), aiming to influence the equilibrium price. Local generators or \emph{followers} can only act after observing the leader's strategy. Note here that the terms line investor and leader (player 1) are used interchangeably, as are local generators and follower (player 2). The equilibrium of the game is found by \emph{backward induction}. The line investor estimates the best response of local generators, given their own generation capacity, and then decides a strategy that maximises their profit. At a second stage, the follower observes this strategy and decides on their generation capacity to be installed, given by the best response function, i.e. maximising their own profit, as anticipated by the leader.

\subsection{Stackelberg game with stochastic generation and varying demand}
\label{Stack_Model}

This section presents the formal solution of the network upgrade problem with stochastic generation. Crucially, the equilibrium of the game depends on the curtailment imposed to the players. However, curtailment depends on the wind resource at the project location and varying demand. To ease understanding, we first formulate the problem for a single renewable player, then we expand this to a two-player setting.

\subsubsection{Single player analysis}
\label{Single_player}
We define as $x_i$ the per unit or normalised power generated by a single generator type $i$ (e.g. player 1 or line investor) at a certain location without curtailment. Essentially, $x_i$ is a stochastic variable which depends on the wind speed distribution and is equal to $x_i={P_{G_i}}/{P_{N_i}}$, where $P_{G_i}$ is the actual power output of $i$ generator and $P_{N_i}$ the rated capacity. By definition, $x_i$ is bounded in the region $x_i\in[0,1]$. Moreover, we assume the normalised power generated follows a probability distribution function $f(x_i)$, such that$\displaystyle{\int_0^1f(x_i)dx_i=1}$. The expected power generated when no curtailment is assumed, is equal to:
\begin{equation}
\mathbb{E}(P_{G_i})=\mathbb{E}(x_i)\cdot P_{N_i}
=\int_0^1{x_i}P_{N_i}f(x_i)dx_i
\label{Eq:EGi}
\end{equation}
At areas with network constraints, generators are often curtailed. In this case, if the power demanded (at location $A$) at each time interval $t$ is equal to $P_{D,t}$ (we can safely assume that the power demanded is well known and predicted), then the expected curtailed power or expected curtailment is required if and only if there is excess generation $P_{G_i}-P_{D,t}>0$ resulting to $x_i>{P_{D,t}}/{P_{N_i}}$. Time step $t$ can be defined as a reasonable time step, e.g. one hour or so that it coincides with the resolution of available wind speed data or demand data. The expected power curtailed (for time interval $t$) is the difference between the conditional expectation of the power generated minus the demand under the condition that it exceeds that demand, multiplied by the probability that the power generated exceeds the demand. In other words, the expected curtailment is equal to the expected value of the generation given that generation exceeds the demand (a posteriori expectation) minus the power demanded times the probability that generation exceeds the demand:
\begin{equation}
\mathbb{E}({P_{C_i,t}})=[\mathbb{E}(P_{N_i}\cdot x_i\mid_{P_{N_i}\cdot x_i>P_{D,t}})-P_{D,t}]\cdot \displaystyle{ \int_{\frac{P_{D,t}}{P_{N_i}}}^1f(x_i)dx_i}
\end{equation}
The first term (conditional expectation) is by definition  equal to:
\begin{equation}
\mathbb{E}(P_{N_i}\cdot x_i\mid_{P_{N_i}\cdot x_i>P_{D,t}})=\frac{\displaystyle{\int_{ \frac{P_{D,t}}{P_{N_i}}}^1P_{N_i}x_if(x_i)dx_i}}{\displaystyle{\int_{\frac{P_{D,t}}{P_{N_i}}}^1f(x_i)dx_i}}
\end{equation}
Therefore, the expected curtailment is given by:
\begin{equation}
\mathbb{E}({P_{C_i,t}})=\int_{ \frac{P_{D,t}}{P_{N_i}}}^1P_{N_i}x_if(x_i)dx_i-P_{D,t}\int_{\frac{P_{D,t}}{P_{N_i}}}^1f(x_i)dx_i
\end{equation}

\subsubsection{Two player analysis}
Following the same intuition as in Section~\ref{Single_player}, we can estimate the expectations of power produced and curtailed for a two-player game. There are two types of generators, leader (player 1) and follower (player 2), both located at area B, but at different sub-regions of this location, experiencing different but correlated wind and satisfying the same aggregate demand. The stochastic variables $x_1$ and $x_2$, follow a joint probability distribution function  $f(x_1,x_2)$, which satisfies the property $\displaystyle{\int_0^1 \int_0^1 f(x_1,x_2)dx_2dx_1=1}$. Note here that a joint probability distribution is assumed as in practice the wind resources of the players are likely to be correlated, e.g. neighbouring wind farms experience high winds at the same time and so on. The total expected power generated without any curtailment is equal to:
\begin{equation}
\mathbb{E}(P_{G})=\int_0^1 \int_0^1 (x_1P_{N_1}+x_2P_{N_2})f(x_1,x_2)dx_2dx_1
\label{Tot_G}
\end{equation}
Similarly to the analysis in Section~\ref{Single_player}, the power is curtailed when there is excess generation or $x_1P_{N_1}+x_2P_{N_2}-P_{D,t}>0$, resulting to $\displaystyle{x_2>\frac{P_{D,t}-x_1P_{N_1}}{P_{N_2}}}$. Expected curtailment is equal to the expected value of generation given that generation exceeds the demand (a posteriori expectation) minus the power demanded times the probability that generation exceeds the demand:
\begin{equation}
\mathbb{E}({P_{C,t}})=[\mathbb{E}(x_1P_{N_1}+x_2P_{N_2}\mid_{x_1P_{N_1}+x_2P_{N_2}>P_{D,t}})-P_{D,t}]\cdot \int_0^1 \int_{\frac{P_{D,t}-x_1P_{N_1}}{P_{N_2}}}^1 f(x_1,x_2)dx_2dx_1
\end{equation}
The conditional expectation of the generation given that a curtailment event has happened (i.e. generation exceeds the demand) is by definition equal to:
\begin{equation}
\mathbb{E} (x_1P_{N_1}+x_2P_{N_2}\mid_{x_1P_{N_1}+x_2P_{N_2}>P_{D,t}})=\frac{\displaystyle{\int_0^1\int_{\frac{P_{D,t}-x_1P_{N_1}}{P_{N_2}}}^1(x_1P_{N_1}+x_2P_{N_2})f(x_1,x_2)dx_2dx_1}}{\displaystyle{\int_0^1\int_{\frac{P_{D,t}-x_1P_{N_1}}{P_{N_2}}}^1f(x_1,x_2)dx_2dx_1}}
\end{equation}
Combining the last two equations, the expected curtailment at time interval $t$ is:
\begin{equation}
\mathbb{E}({P_{C,t}})=\int_0^1\int_{\frac{P_{D,t}-x_1P_{N_1}}{P_{N_2}}}^1(x_1P_{N_1}+x_2P_{N_2})f(x_1,x_2)dx_2dx_1 - P_{D,t}\int_0^1\int_{\frac{P_{D,t}-x_1P_{N_1}}{P_{N_2}}}^1f(x_1,x_2)dx_2dx_1
\label{Tot_C}
\end{equation}

Equations~(\ref{Tot_G}) and (\ref{Tot_C}) are the derived expressions for the expected power generated and curtailed at each time step $t$, respectively. For a longer period of time, e.g. equal to the project lifetime, the total energy produced by both players is $\displaystyle{E_{G}=\sum_t \mathbb{E} (P_{G_{t}}), \forall t}$, as derived by (\ref{Tot_G}). In a similar fashion, by Eq.~(\ref{Tot_C}) the energy curtailed by both players is $\displaystyle{E_{C}=\sum_t \mathbb{E} (P_{C_{t}}), \forall t}$. However, the profit equations in Eq.~(\ref{Prof_1}) and  Eq.~(\ref{Prof_2}) require expressions of the expected energy generated and curtailed by each player. The energy that could have been produced by $i$ player (if no curtailment is imposed) is $\displaystyle{E_{G_i}=\sum_t x_iP_{N_i}=\sum_t \mathbb{E}(P_{G_{i,t}}), \forall t}$ as in Eq.~(\ref{Eq:EGi}) and the energy curtailed for each $i$ player is defined as $\displaystyle{E_{C_i}=\sum_t P_{C_{i,t}}, \forall t}$. Individual curtailment for `fair' curtailment rules (such as Pro Rata or FRR) is shared amongst the generators proportionally to their size or actual power output at the time of curtailment. For instance, if at time interval $t$ there is excess generation and $P_{C,t}$ is to be curtailed then the power curtailed by each player $i$ is equal to:
\begin{equation}
P_{C_{i,t}}=\frac{x_{i,t}P_{N_i}}{x_{i,t}P_{N_i}+x_{-i,t}P_{N_{-i}}}\cdot P_{C,t}
\label{ProRata}
\end{equation}
where $-i$ denotes all other players. As shown in Section~\ref{Illustration}, `fair' rules lead to approximately equal CF reduction, in the long term. Therefore, the energy curtailed by each player throughout the project lifetime can be approximated based on Eq.~(\ref{ProRata}) as:
\begin{equation}
E_{C_i}=\frac{E_{G_i}}{E_{G_i}+E_{G_{-i}}}E_C
\end{equation}

In the above expressions, the profits as defined in Eq.~(\ref{Prof_1}) and Eq.~(\ref{Prof_2}) are functions of the players' strategies, i.e. the rated capacity they install. In particular, $E_{G_i}$ is a function of $P_{N_i}$, but $E_{C_i}$ is a function of both players' rated capacities $P_{N_1},P_{N_2}$. Therefore, the research question can be rephrased as \textit{`Which are the optimal rated capacities  players install at the equilibrium of the game, so that profits are maximised?'} As players do not have the same market power, initially the leader defines a set of feasible solutions to their control variable $P_{N_1}$. 
Given the generation capacity installed by the line investor $P_{N_1}$, the local generators best response is:
\begin{equation}
P_{N_2}^{*}= \argmax_{P_{N_2}} {\Pi_2(P_{N_1},P_{N_2}})
\label{BestResp1}
\end{equation}
Next the leader estimates which solution from the set of the follower's best response maximises their profits. Given the capacity built by the followers $P_{N_2}^{*}$, the line investor's best response is:
\begin{equation}
P_{N_1}^{*}= \argmax_{P_{N_1}}\Pi_1(P_{N_1},P_{N_2}^{*})
\label{BestResp2}
\end{equation}
In other words, the leader moves first by installing their own generation capacity. In the second level, followers respond to the generation capacity built, as anticipated by the leader. The equilibrium of the game  $(P_{N_1}^*,P_{N_2}^*)$ satisfies both Eq.~(\ref{BestResp1}) and Eq.~(\ref{BestResp2}) and is given by the notion of the subgame perfect equilibrium.

In the following section, we show how the methodology is applied in practice using historical real data available for accurate representation.

\section{Application of grid reinforcement}
\label{Cas_Stud}
In this section, we apply the theoretical results of our analysis to a real network upgrade problem in the UK, namely the link between Hunterston and Kintyre in Western Scotland.
\begin{figure}
\centering
\includegraphics[width=0.5\textwidth]{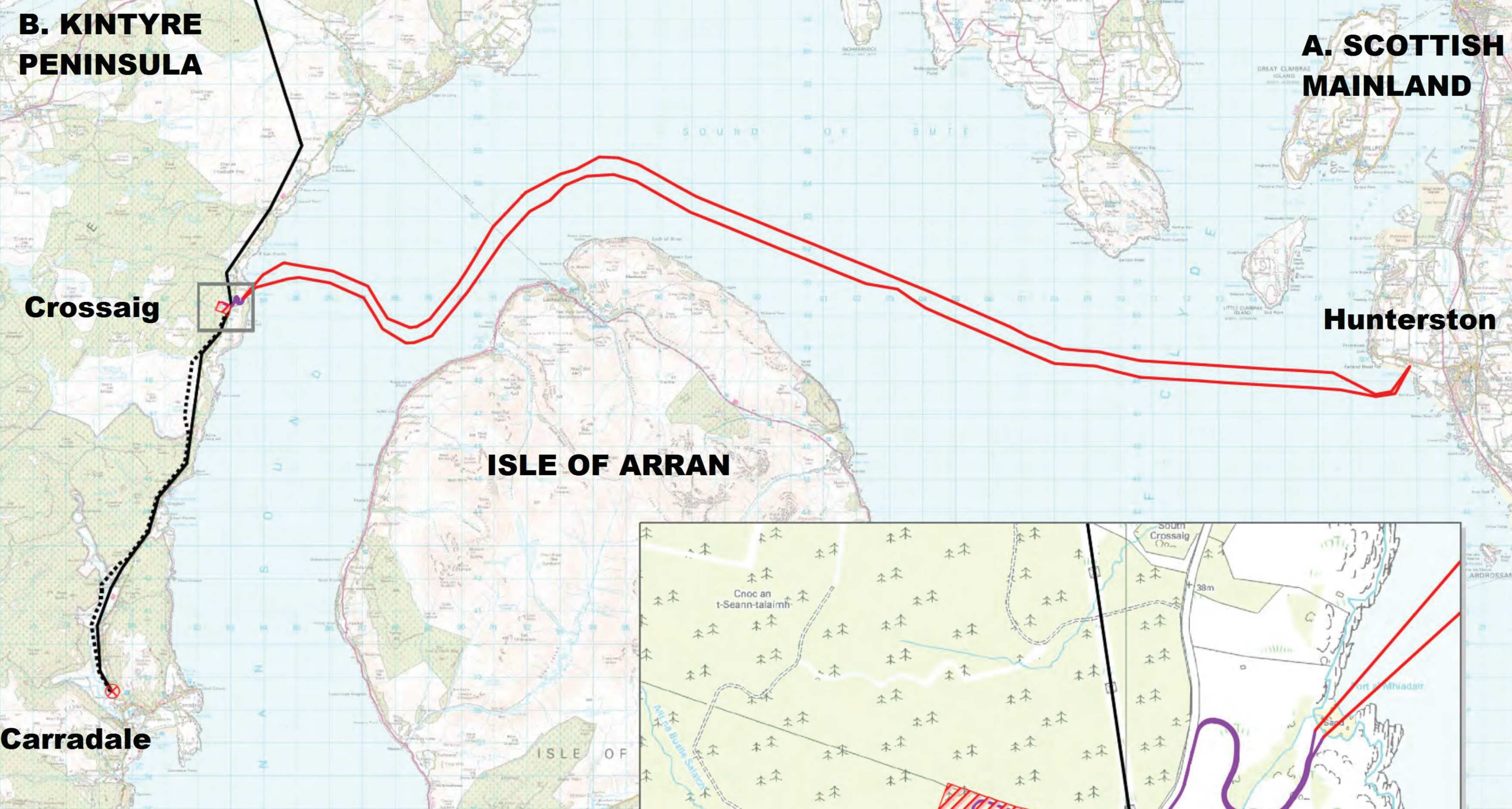}
\caption[Caption]{Kintyre-Hunterston project map: Overhead line from Carradale to Crossaig 13~km, from Crossaig to Hunterston 41~km (subsea cable) and 3.5~km (land cable)\protect \footnotemark}
\label{fig:Kint-Hunt}
\end{figure}

\footnotetext{\url{https://www.ssen-transmission.co.uk/projects/kintyre-hunterston/}}

\subsection{Kintyre-Hunterston link}

The grid reinforcement project links the Kintyre peninsula to the Hunterston substation on the Scottish mainland (see Fig.~\ref{fig:Kint-Hunt}). The project includes the installation of new overhead power lines, a new substation and a double circuit subsea cable of 220kV HVAC placed north of Arran for a distance of 41 km. Kintyre, located in the West of Scotland, is a region that has attracted a vast amount of renewable generation ($454~\text{MW}$ RES capacity was expected to connect by the end of 2015) and high interest in RES investment (more than $793~\text{MW}$ potential connections), predominantly wind generation. In fact, the growth of renewable generation in the Kintyre region was responsible for the growing stress in the existing transmission line, originally designed and built to serve a typical rural area of low demand. According to SSE (the DNO in this region), the Kintyre-Hunterston project will provide $150~\text{MW}$ of additional renewable capacity and it will cost $\pounds 230\rm{m}$. Apart from facilitating renewable generation, the project is expected to increase security of supply and export capability to the mainland grid, delivering value to consumers estimated at $\pounds 18\rm{m}$ per annum \cite{SKM04}.

\subsection{Problem setting}
Based on the figures of this project, we apply the methodology described in Section~\ref{Net_up}
and characterise the equilibrium of the Stackelberg game. We assume that the demand region or Location A is Hunterston and location B is the geographical region covering the Kintyre peninsula. The line investor installs their own generation capacity at a sub-region of location B and local generators install wind capacity at a different sub-region of location B. The same notation applies here as the line investor represents player 1 (leader) and local generators represent player 2 (follower).

\subsubsection{Wind speed data}

We use real wind speed data to perform our analysis, provided by the UK Met Office (UK Midas Dataset)\footnote{\url{https://badc.nerc.ac.uk/search/midas_stations/}}. To model the two sub-regions at location B, two representative MIDAS weather stations were selected, the weather station with ID 908 located in the Kintyre peninsula and with ID 23417 located in Islay, with a distance between them of $44~\rm{km}$. The two stations were selected for a variety of reasons such as, the time period of available data and their proximity so that the wind speed correlation between neighbouring locations could be modelled. We will use station 908 as the location of the wind farm of the line investor and station 23417 for the local generators wind farm.

The weather stations provide hourly measurements, in particular the hourly average of the mean wind speed (hourly measurements of a 10-min averaged wind speed), as measured at anemometer's height, rounded to the nearest knot ($1~\text{kn}=0.5144~\text{m/s}$). The nominal anemometer heights are $z_{a}=10~\text{m}$, for the  leader and follower's location. Any missing data of a shorter duration than $6~\text{hr}$ were replaced by a linear interpolation of the nearest available wind speeds, rounded to the nearest knot. This technique does not introduce a large error, as it is unusual to have a large variation of the wind speed for such a short time period and the data substituted represent a small percentage of total measurements available.

\begin{figure}
\centering
\includegraphics[width=0.5\textwidth]{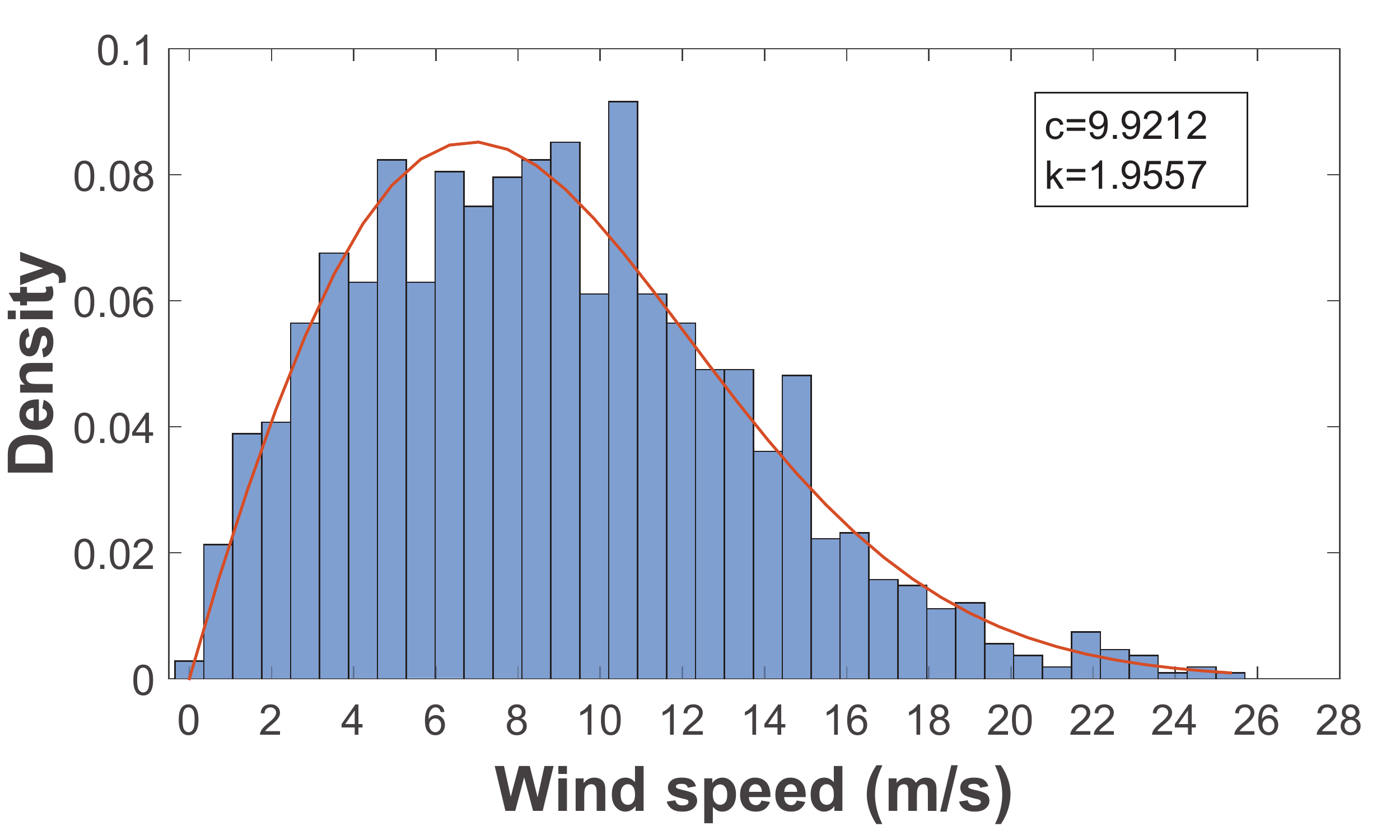}
\caption{Wind speed histogram and best fit Weibull curve (local generators' location)}
\label{fig:Weib}
\end{figure}

The analysis refers to the period 1999-2015, as this is the time frame when the stations have data in common. This 17-year period of examination is approximately equal to the typical lifetime of a renewable generation project (20 years). Given the hourly wind speeds, we can estimate the power output generated by a typical wind turbine. We use a generic power curve based on an Enercon E82 wind turbine\footnote{\url{http://www.enercon.de/en/products/ep-2/e-82/}} of $2.05~\text{MW}$ rated capacity and hub height of $85~\text{m}$. The wind turbine has a cut-in wind speed of $3~\text{m/s}$, a cut-out wind speed of $28~\text{m/s}$, and a rated wind speed of $13~\text{m/s}$ at which the turbine generates rated power output. The wind speeds need to be extrapolated to hub height. We use a logarithmic shear profile to estimate the wind speed at hub height $u_h$:
\begin{equation}
u_h=u_a\frac{\log\nicefrac{\displaystyle{z_h}}{\displaystyle{z_o}}}{\log\nicefrac{\displaystyle{z_a}}{\displaystyle{z_o}}}
\end{equation} 
where $u_a$ is the wind speed at anemometer height, $z_a=10~\text{m}$ the anemometer height, $z_h=85~\text{m}$ the hub height and $z_o=30~\text{mm}$ is the surface roughness (short grass), which represents a typical environment for the weather stations and is also used in \cite{fruh2015local,watson2015wind}.

The literature in wind forecasting \cite{tuller84characteristics, edwards2001level} commonly uses Weibull distributions for the representation of actual wind distributions, and we adopt this methodology here. For greater accuracy, we need to account for hourly and seasonal changes of the wind speed. For this reason, we consider 96 different Weibull distributions, one for every hour ($1-24$) and season ($1-4$). Hour 1 refers to 00:00 and hour 24 to 23:00, March, April and May refer to Spring, June, July and August to Summer and so on. This approach is required to associate power generation caused by wind conditions to the power demanded, which also depends highly on the time of day and season, as shown in Section~\ref{dem_data}. Each probability distribution function is approximated by a Weibull function with a shape $k$ and scale factor $c$:
\begin{equation}
f(u,c,k)=\frac{k}{c}\left( \frac{u}{c}\right) ^{(k-1)}e^{-\displaystyle{\left( \frac{u}{c}\right) }^k}
\end{equation}

The parameters of the Weibull distributions are found by means of the function `fitdist' in MATLAB. For example, Fig.~\ref{fig:Weib} shows the wind speed histogram and the best Weibull fit for 09:00 hour in Autumn.

\begin{figure}
\centering
\includegraphics[width=0.5\textwidth]{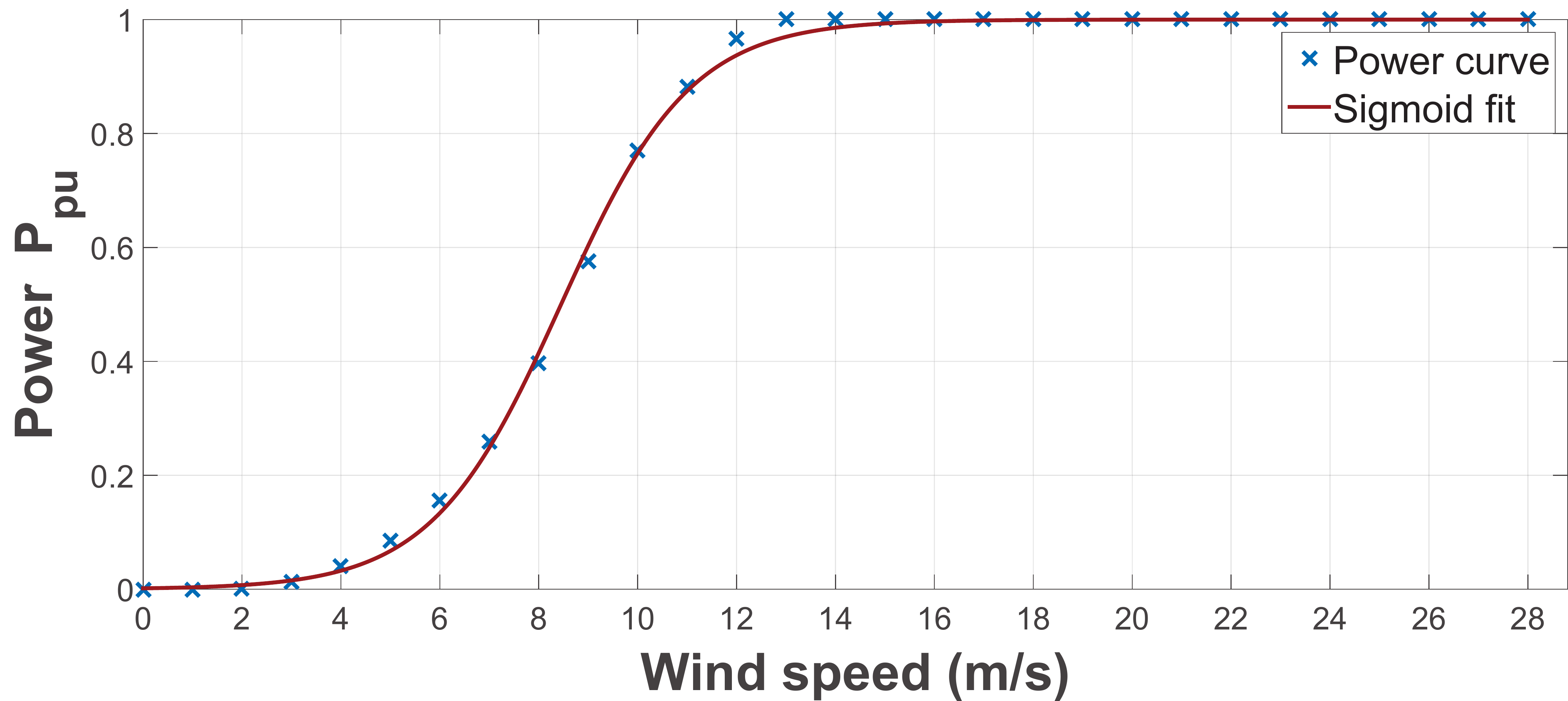}
\caption{Power curve of Enercon E82 and best Sigmoid fit function}
\label{fig:Sigm}
\end{figure}

The power output of a wind turbine is estimated by the power curve given by the manufacturer. The generated power is normalised to the rated capacity or nominal power output $P_{pu}=P/P_{nom}$ and intermediate values are approximated by a sigmoid function with parameters $a=0.3921~\rm{s/m}$ and $b=16.4287~\rm{m/s}$ (see Fig.~\ref{fig:Sigm}):
\begin{equation}
f(u,a,b)=\displaystyle{\frac{1}{1+e^{-a(u-b)}}}
\end{equation}

In many works \cite{fruh2015local, liu2011impact, fabbri2005assessment}, given that the normalised power output is bounded in the closed interval $[0,1]$, the output profile is approximated by a standard beta distribution. Following the same approach we can derive 96 beta probability distribution functions for every hour and season given by:
\begin{equation}
f(x_i)=\frac{1}{B(\alpha,\beta)}x_i^{\alpha-1}(1-x_i)^{\beta-1}
\end{equation}
where
\begin{equation}
B(\alpha,\beta)=\frac{\Gamma(\alpha)\Gamma(\beta)}{\Gamma(\alpha+\beta)} \int_0^1 t^{\alpha-1}(1-t)^{\beta-1}dt
\end{equation}
\vspace{-3mm}
\begin{figure}
\centering
\includegraphics[width=0.5\textwidth]{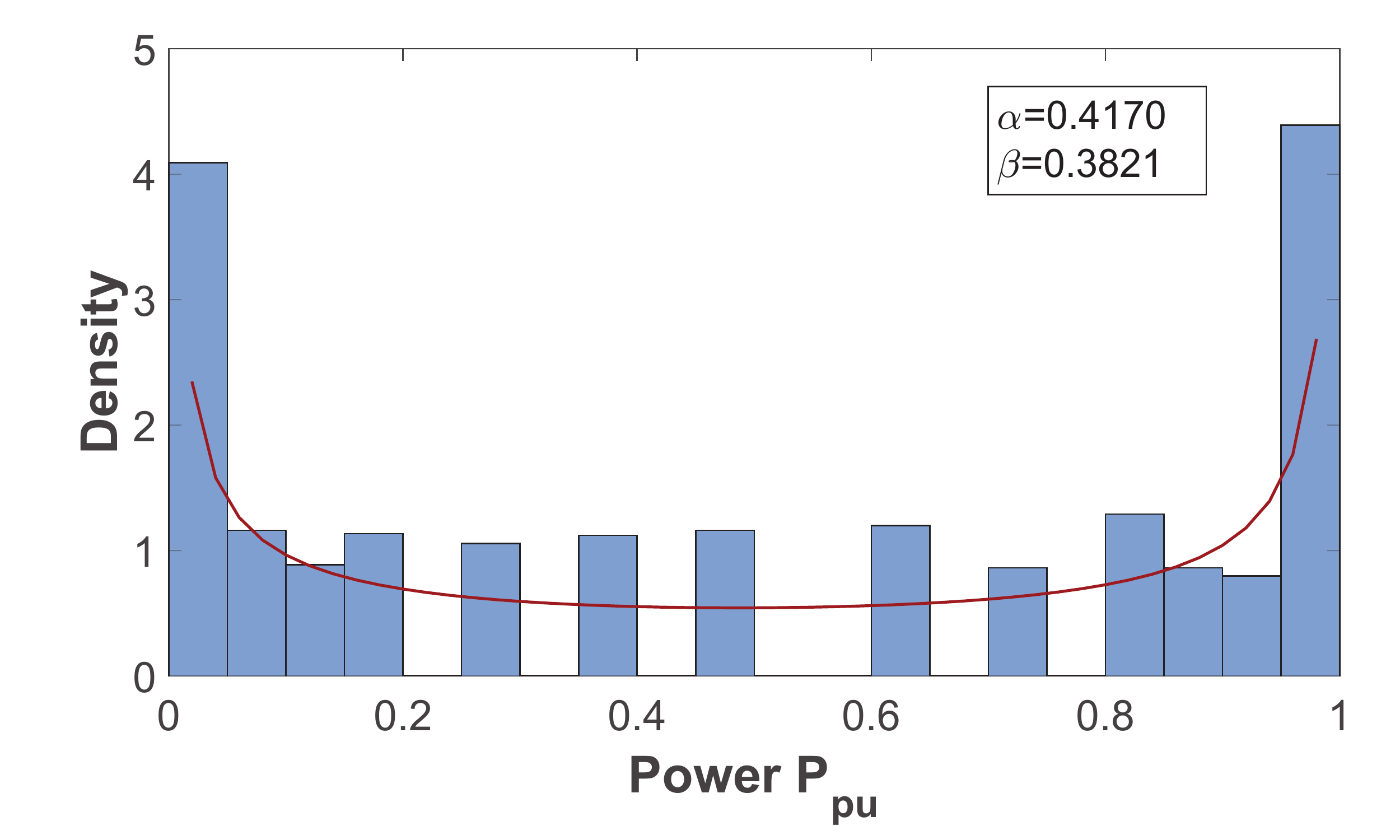}
\caption{Power output histogram and best fit Beta curve (local generators' location)}
\label{fig:Beta}
\end{figure}

Fig.~\ref{fig:Beta} shows the histogram and best beta fit at the line investor's location at 09:00 hour in Autumn. Note here that the effect of empty bins seen in Fig.~\ref{fig:Beta}, is created by the combined effect of the Weibull distribution with the power generated and the fact that the data provided by the UK Met Office dataset are rounded to the nearest knot. 

\begin{figure}
\centering
\includegraphics[width=0.5\textwidth]{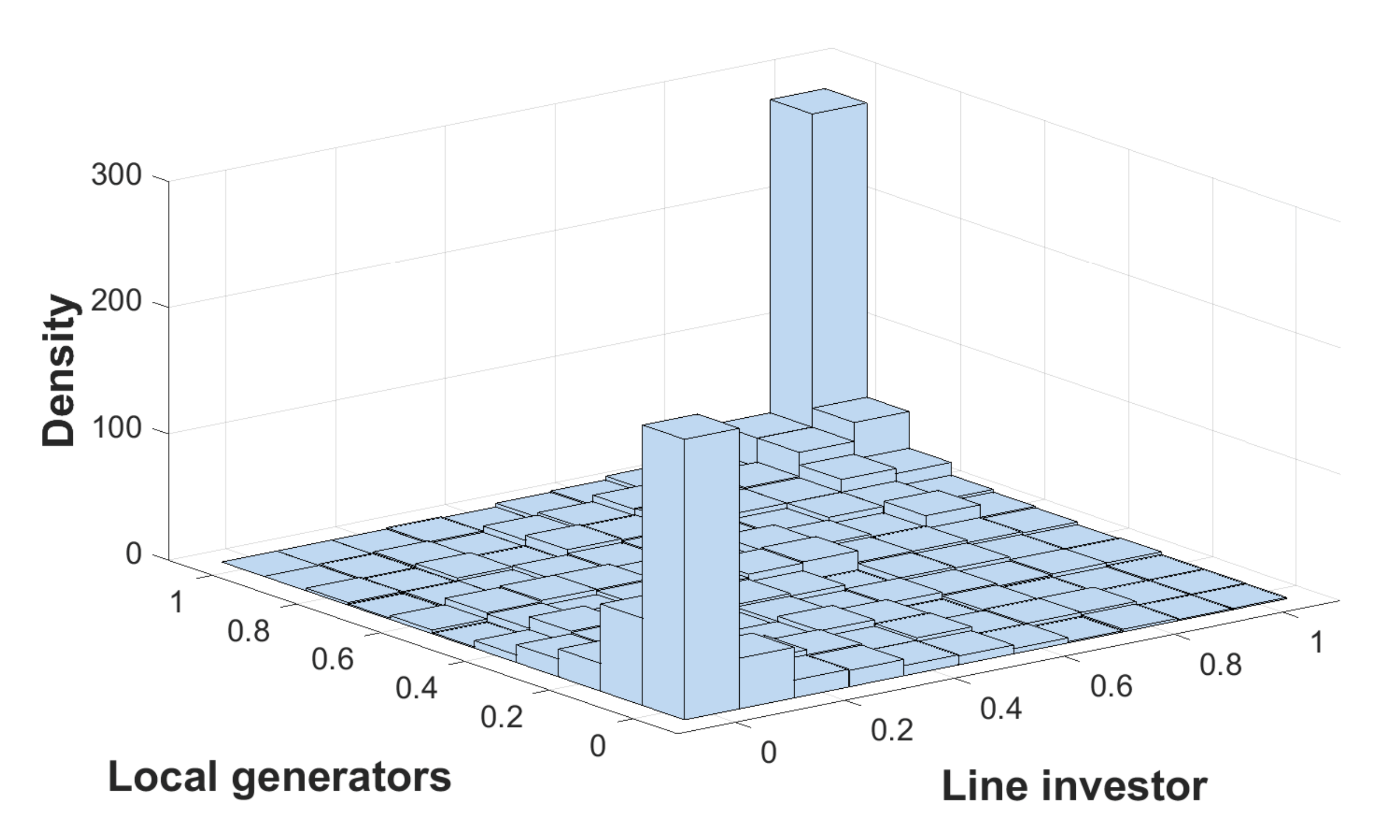}
\caption{Joint power output histogram}
\label{fig:Joint}
\end{figure}

The joint probability distribution of the power outputs of both players takes into account wind speed spatial correlation. Note here that the total power output is not a two-dimensional beta distribution, as the power outputs of the players are correlated. If there are sufficient wind speed measurements for both players locations, then the joint probability distribution can be estimated directly from the available data. For example, Fig.~\ref{fig:Joint} shows the joint power histogram at 09:00 hour in Autumn. Note here that most observations are concentrated at zero power output (no wind) or close to rated power (wind equal to or above nominal). Many observations appear around the diagonal, which indicates partial correlation of the power output generated by the players.

\subsubsection{Demand data}
\label{dem_data}

The demand data used are based on UK National Demand and published by the National Grid (historical demand data)\footnote{\url{http://www2.nationalgrid.com/UK/Industry-information/Electricity-transmission-operational-data/Data-Explorer/}}. The data available range from January 2006 to December 2015 and consist of the national demand in half-hourly intervals, corresponding to the settlement periods of the UK energy market. National demand is estimated as the sum of generation based on National Grid operational metering plus the estimated embedded generation from wind/solar generators plus imports. Real demand may differ from this estimation, as some demand is not visible to the transmission system, due to embedded generation connected to the distribution networks.

Half-hourly demand data were substituted with the hourly average to keep the same resolution as the wind speed data. Next, demand data were analysed in a similar manner to distributions for every hour and season.

\begin{figure*}
\centering
\begin{tabular}{cc}
\begin{subfigure}{0.5\textwidth}
\includegraphics[width=\linewidth]{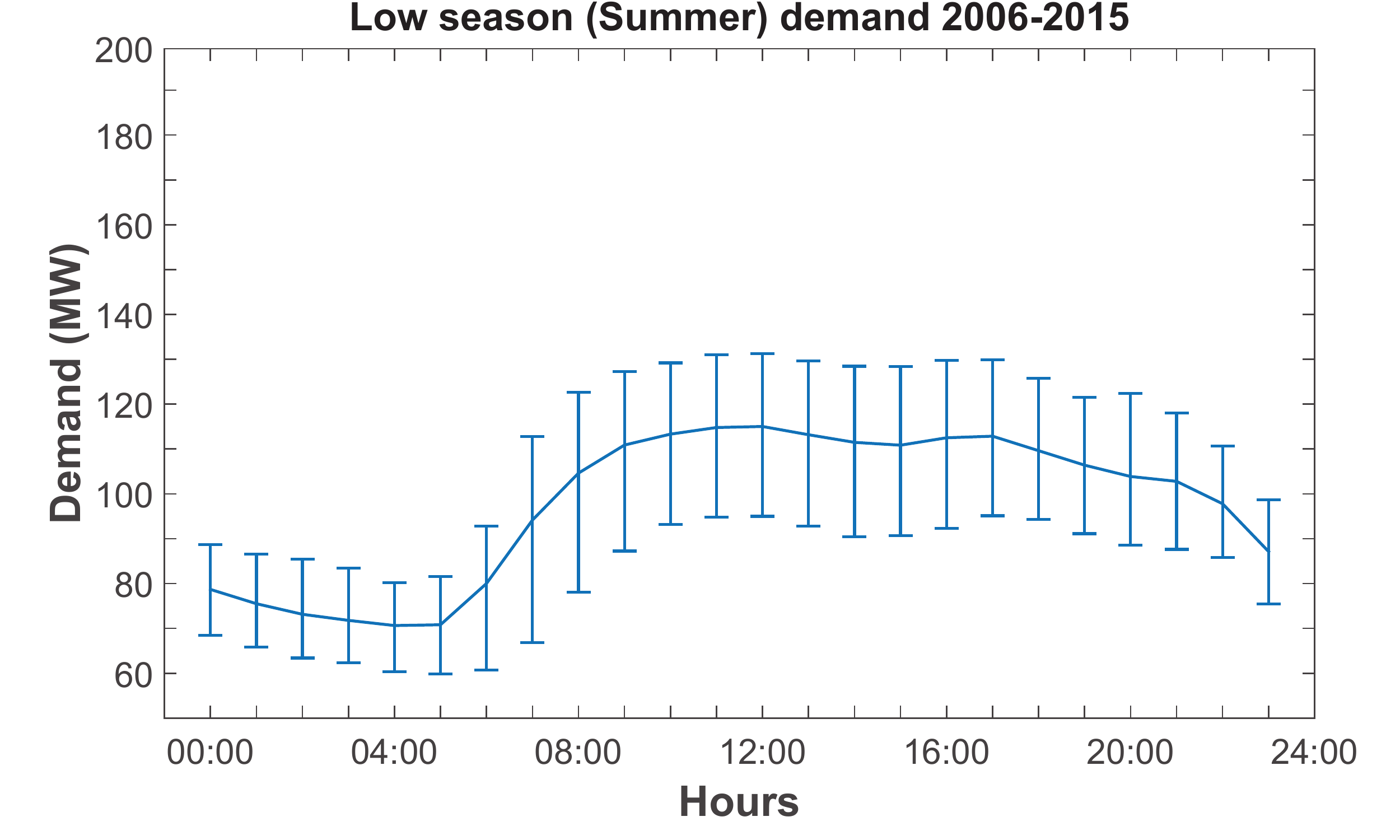} \caption{}
\end{subfigure}
 & 
 \begin{subfigure}{0.5\textwidth}
 \includegraphics[width=\linewidth]{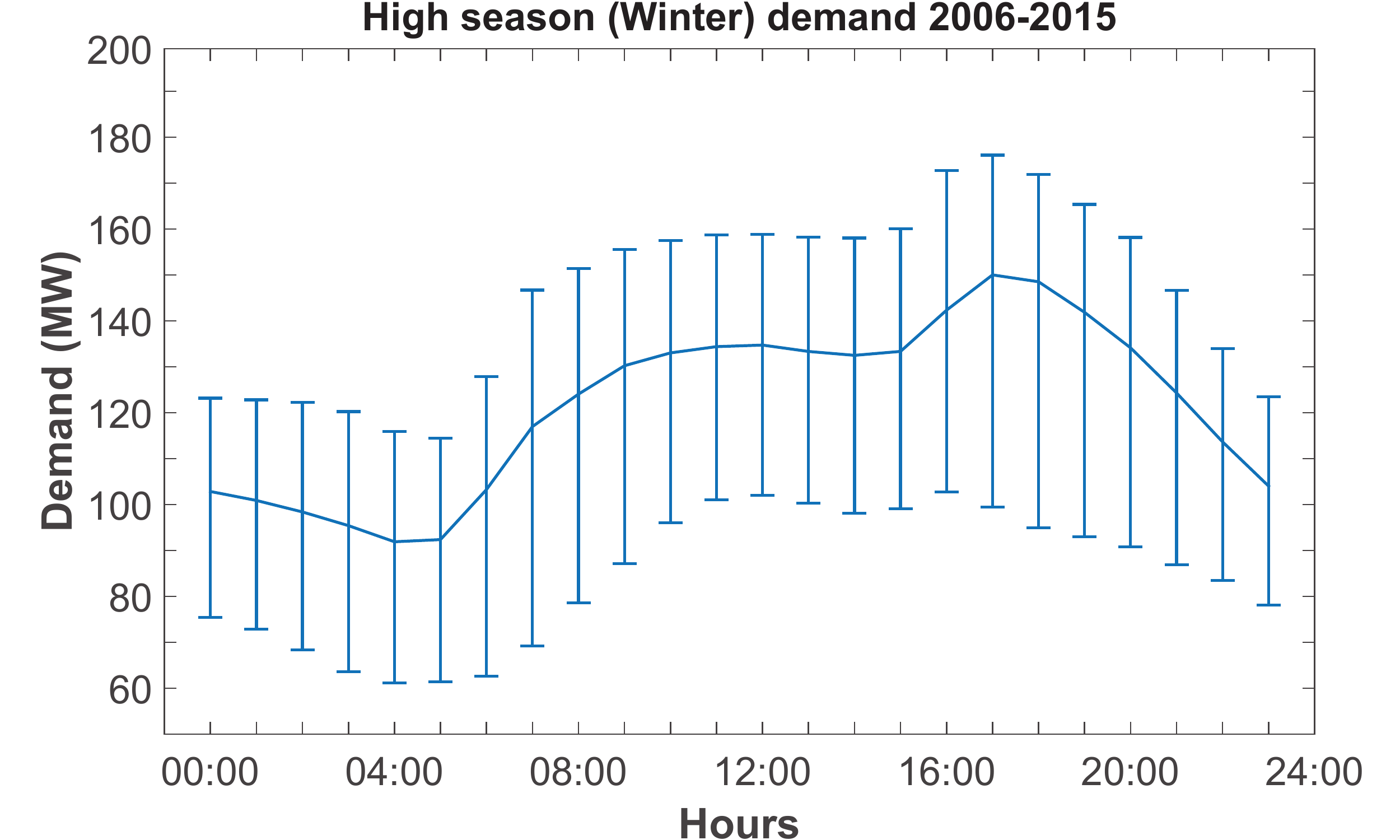} \caption{}
 \end{subfigure}
\end{tabular}
\caption{Seasonal average demand with minimum and maximum values (a) for the lowest demand season (Summer) and (b) uneppeak demand season (Winter)}
\label{Seas_demand}
\end{figure*}

\begin{figure}[h!]
\centering
\includegraphics[width=0.5\textwidth]{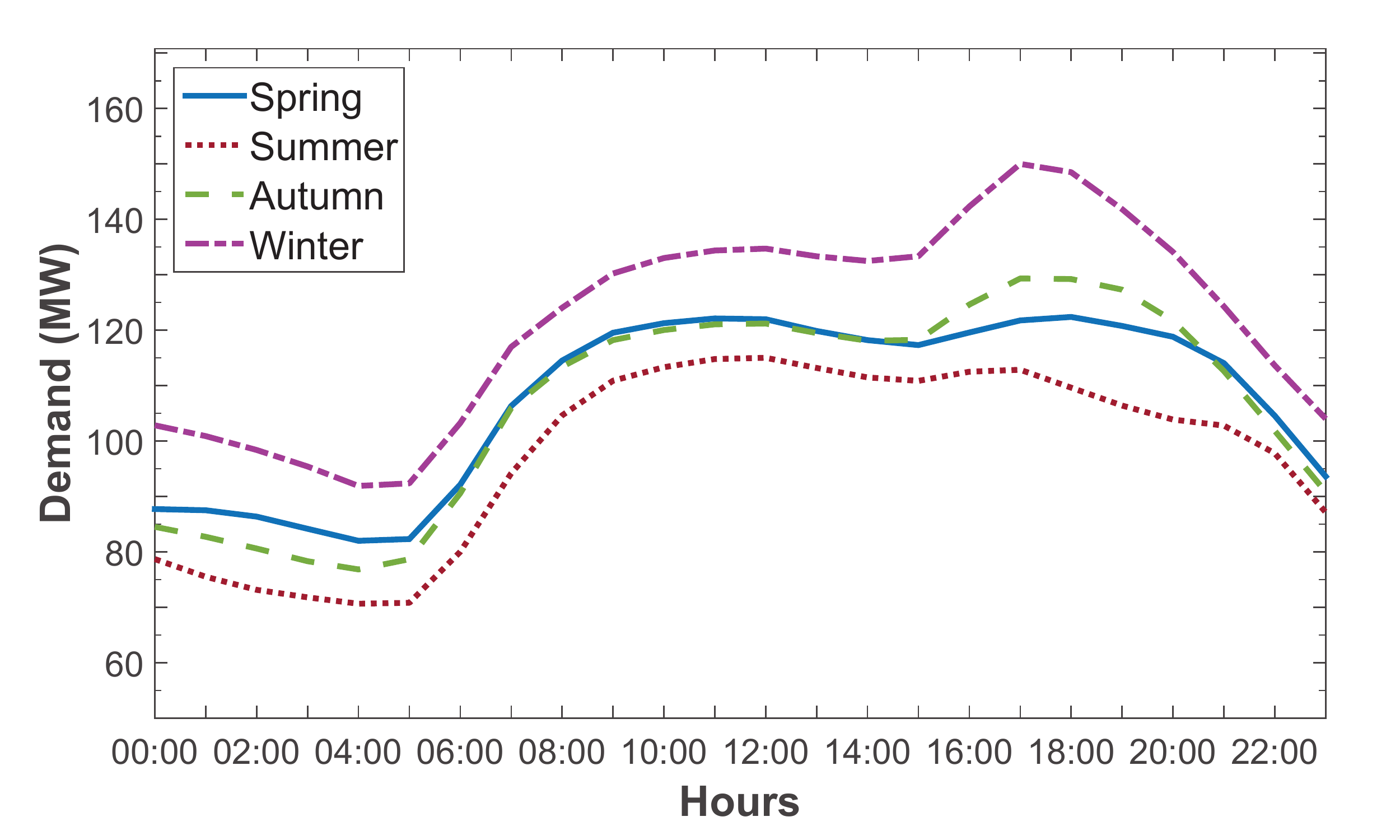}
\caption{Hourly average demand per season}
\label{fig:Seas}
\end{figure}

The demand at location A was taken to be equal to the average national demand for every hour and season, scaled down by a factor, such that peak load is equal to the capacity of the power line, considered equal to $150~\text{MW}$. Indicatively, Fig.~\ref{Seas_demand} shows the average demand at location A together with the minimum and maximum demand for every hour, during the low demand season (Summer) and the high demand season (Winter). In Fig.~\ref{fig:Seas} the seasonal effect on the average demand is shown. The values of average demand per hour and season were used in our experimental setting analysis.

From the demand and joint probability distributions, the equilibrium of the Stackelberg game can be found analytically, as was shown in Section~\ref{Stack_Model}. In the next section, we show a methodology to identify the equilibrium of the game by an empirical approach using actual data from all the hours in a 17 year period.

\subsection{Searching for the empirical game equilibrium using actual data}

This section describes the solution of the game from a data analytical approach. We use the hourly wind speed data for two locations over a period of 17 years and the average demand on an hourly and seasonal basis.

First of all, a feasible solution space needs to be identified. The strategies of the players are the generation capacities they install, meaning the solution space should identify an upper limit of rated capacity, above which any player would only incur losses. All possible solutions are included, even in the case that demand is served by a single player. Based on several trial runs, this upper limit is set to $P_{lim}=415~\text{MW}$ for each player. Note here that $P_{lim}$ is larger than the peak demand ($150~\text{MW}$) divided by the minimum CF experienced by any player at all hours and seasons, to guarantee that all possible solutions are included in the search space. Moreover, the incremental generation capacity a generator can install is set to $0.5~\text{MW}$, therefore the solution search space is defined as $[P_{N_1},P_{N_2}]=[0:0.5:415, 0:0.5:415]$.

For every possible combination of the rated capacities installed $(P_{N_1},P_{N_2})$, we estimate the power generated and curtailed for both players under a fair curtailment rule such as Pro Rata or FRR. For example, for a particular combination of $(P_{N_1},P_{N_2})$, we estimate the power generated at each hour given the wind speed and estimate the power curtailed given the demand. Next, we estimate the aggregate power generated and curtailed by each player for the time period of $17$ years and therefore derive the energy that would have been generated (if no curtailment) and the energy curtailed, as the summation of $145077$ valid data points (hours in the $17$ years that wind speed and demand data are available). For given cost parameters $(c_{G_1},c_{G_2},p_T)$, the profits of both players are estimated by Eq.~(\ref{Prof_1}) and Eq.~(\ref{Prof_2}). Essentially, for every scenario or given $(c_{G_1},c_{G_2},p_T)$, profits are derived for all feasible solutions included in the search space.

The Stackelberg equilibrium is found as follows. Given a certain rated capacity built by the line investor $P_{N_1}$, we find the rated capacity $P_{N_2}^*$ that maximises the follower's profits i.e. $\Pi_2^*$. This step finds the best response of the follower given the strategy of the leader. 
Note here that this results in a solution vector $P_{N_2}^*$, for every $P_{N_1}=[0:0.5:415~\text{MW}]$. Next, from this set of solutions (follower's best response), the leader finds the solution that maximises their own profit i.e. $P_{N_1}^*$, by searching the normal form matrix of the corresponding Stackelberg game. The equilibrium of the game is given by the pair $(P_{N_1}^*,P_{N_2}^*)$, which satisfies both response functions of the two players.

In the next section, we provide the solutions of the empirical study for varying parameters and discuss the main findings of this approach.

\section{Empirical results}
\label{Res}

We assume different scenarios to examine how the equilibrium results depend on varying parameters. Scenario 1 shows the dependence on local generators' cost, Scenario 2 on line investor's cost and Scenario 3 on the the transmission fee. At each scenario, the key parameter varies, while other parameters remain fixed.

In all scenarios, the energy selling price is set equal to $p_G=\pounds 74.3 /\text{MWh}$ (equivalent to a medium sized wind turbine with feed-in tariff and export fee of $\pounds2.52\text{p/}\text{kWh}$ and $\pounds 4.91\text{p/}\text{kWh}$ respectively)\footnote{\url{https://www.ofgem.gov.uk/publications-and-updates/feed-tariff-fit-tariff-table-1-april-2016-non-pv-only}}. All parameters are expressed as a percentage of the $p_G$ for easier representation of the results. The step size of varying parameters is set equal to $0.02p_G$ for all scenarios considered.

Recall here, the follower will install generation capacity $P_{N_2}$ as long as the revenues earned, depending on $p_G-p_T$, are larger than the cost of installing this capacity, depending on $c_{G_2}$. Crucially, the revenues depend on the curtailment imposed to the local generators $E_{C_2}$, which is interdependent on the capacity installed by the line investor $P_{N_1}$. On the other hand, the line investor earns revenues from the energy generated $E_{1}(P_{N_1})$, depending on $p_G$ and the energy transported through the line, $E_2(P_{N_2})$, which is charged with $p_T$. Both $E_1$ and $E_2$ depend on the curtailment imposed, which is a function of the rated capacities installed $E_C(P_{N_1},P_{N_2})$. The costs associated include the generation cost $c_{G_1}$ and the fixed cost of the line $C_T$. The line investor will install generation capacity himself as long as the cost of installing an additional generation unit results in increasing the profit.

\subsection{Scenarios results}

\begin{itemize}
\item \textit{Scenario 1: Varying local generators' cost:} In this scenario, the line investor's generation cost is $c_{G_1}=0.30p_G$ and the transmission fee is $p_T=0.26p_G$, while the local generators' cost varies from $c_{G_2}=0.06p_G$ to $0.52p_G$. Results are shown in the first column of Fig.~\ref{Stackel_graphs}.

\begin{figure*}
\centering
\begin{tabular}{|p{0.01cm}c|c|c|}
\hline
\small(1) &
\begin{subfigure}{0.33\textwidth}
\includegraphics[width=\linewidth]{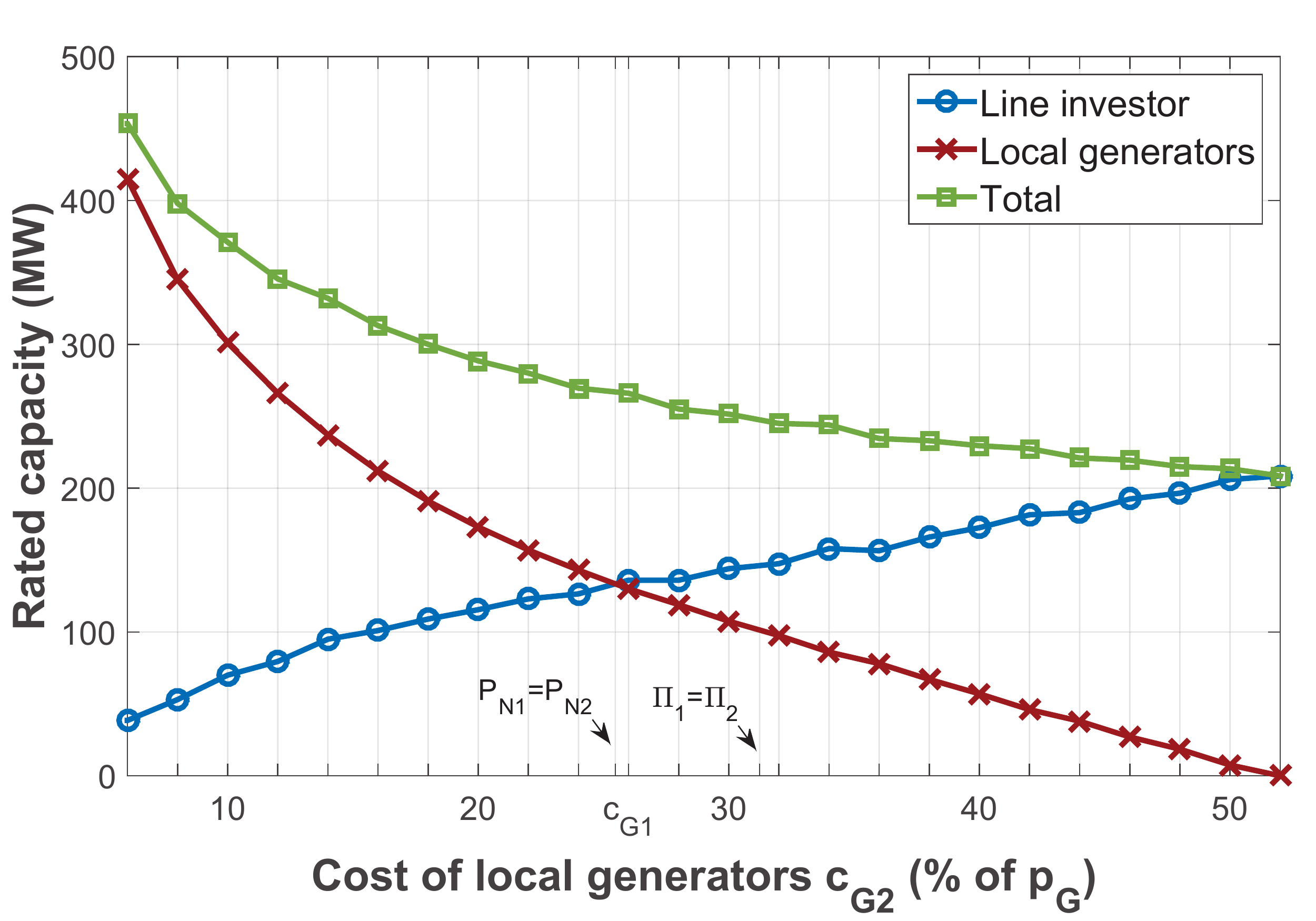}
\end{subfigure}
 & 
 \begin{subfigure}{0.33\textwidth}
 \includegraphics[width=\linewidth]{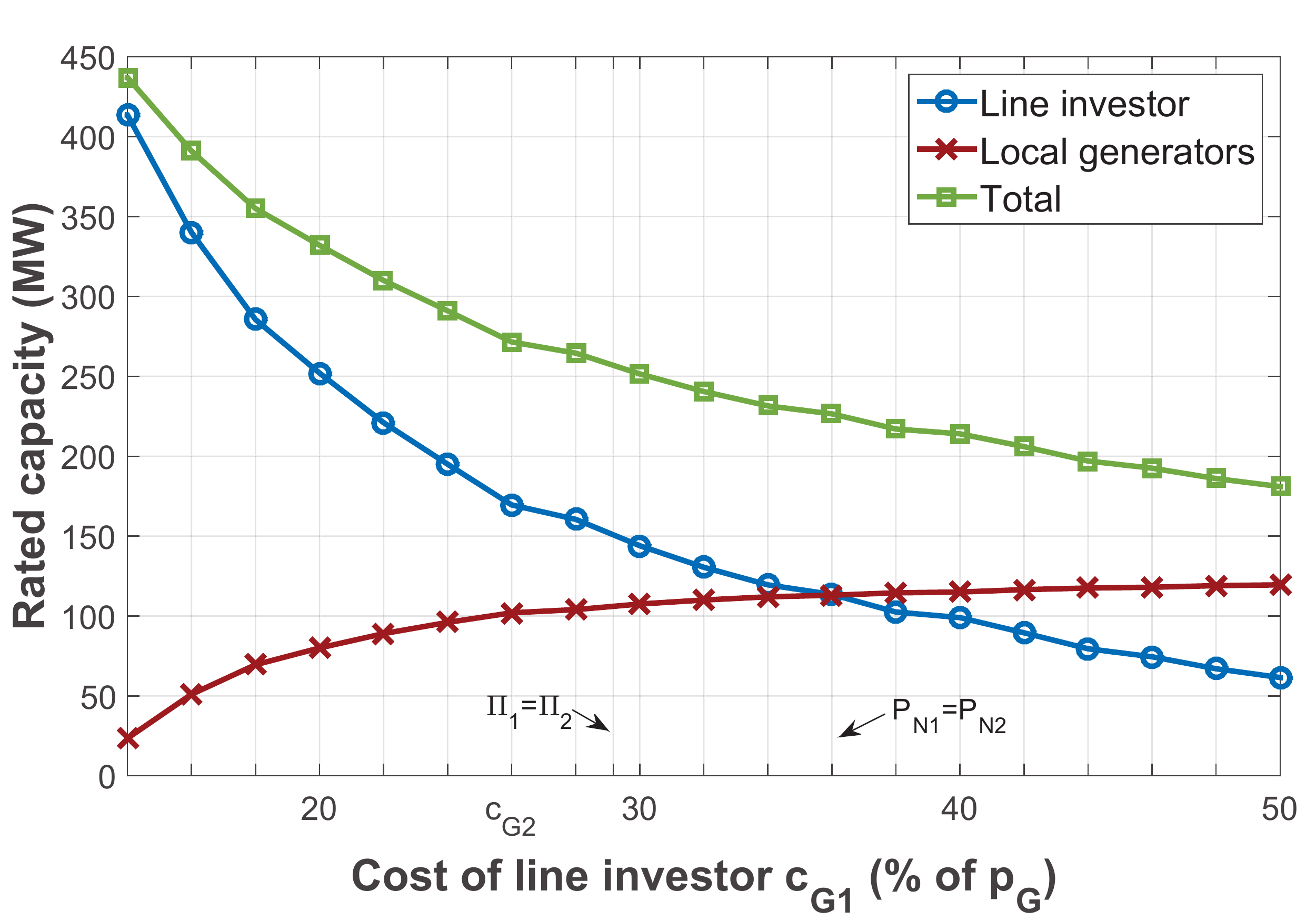}
 \end{subfigure} &
 \begin{subfigure}{0.33\textwidth}
 \includegraphics[width=\linewidth]{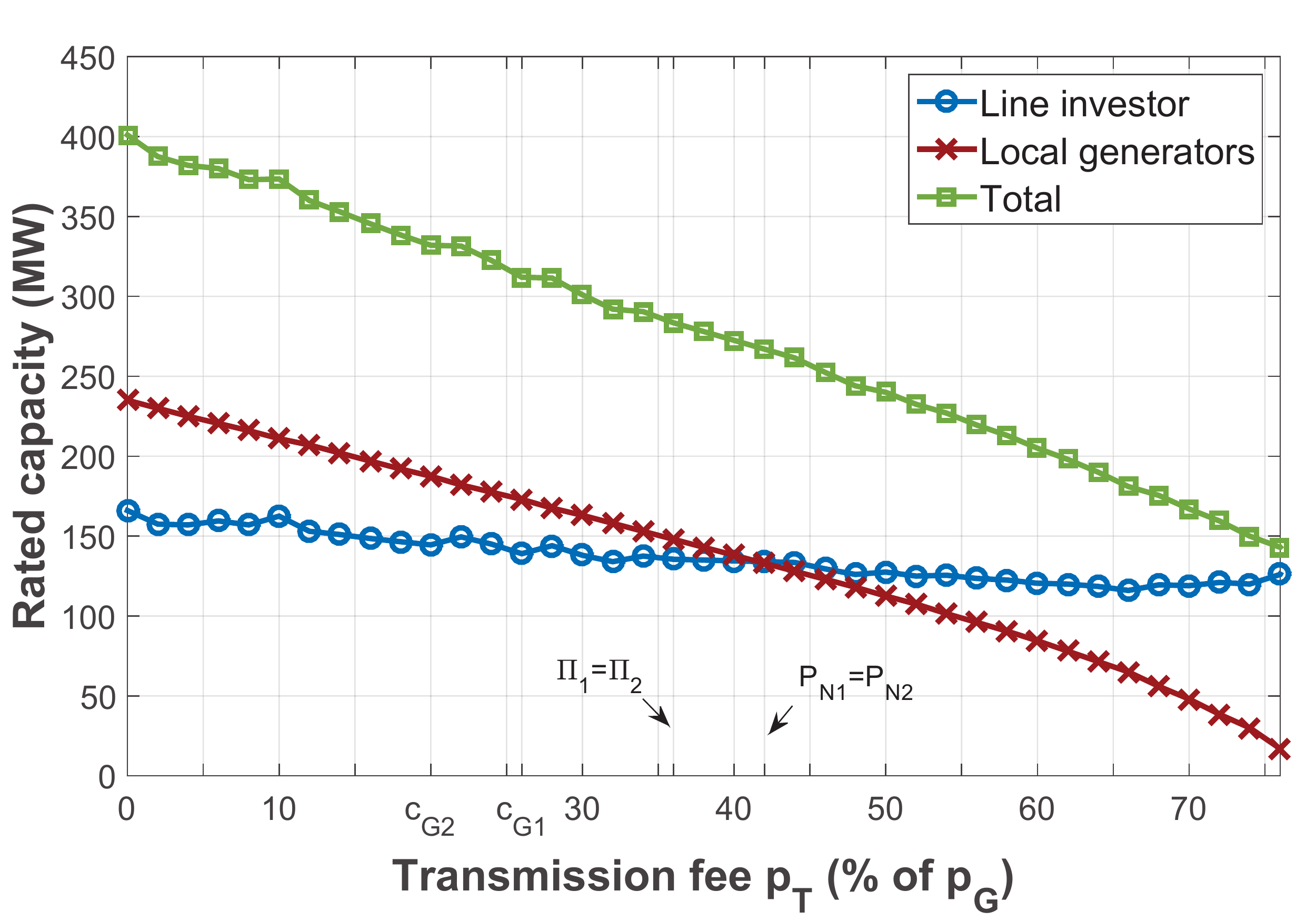}
 \end{subfigure}
 \\
 \small(2) &\begin{subfigure}{0.33\textwidth}
\includegraphics[width=\linewidth]{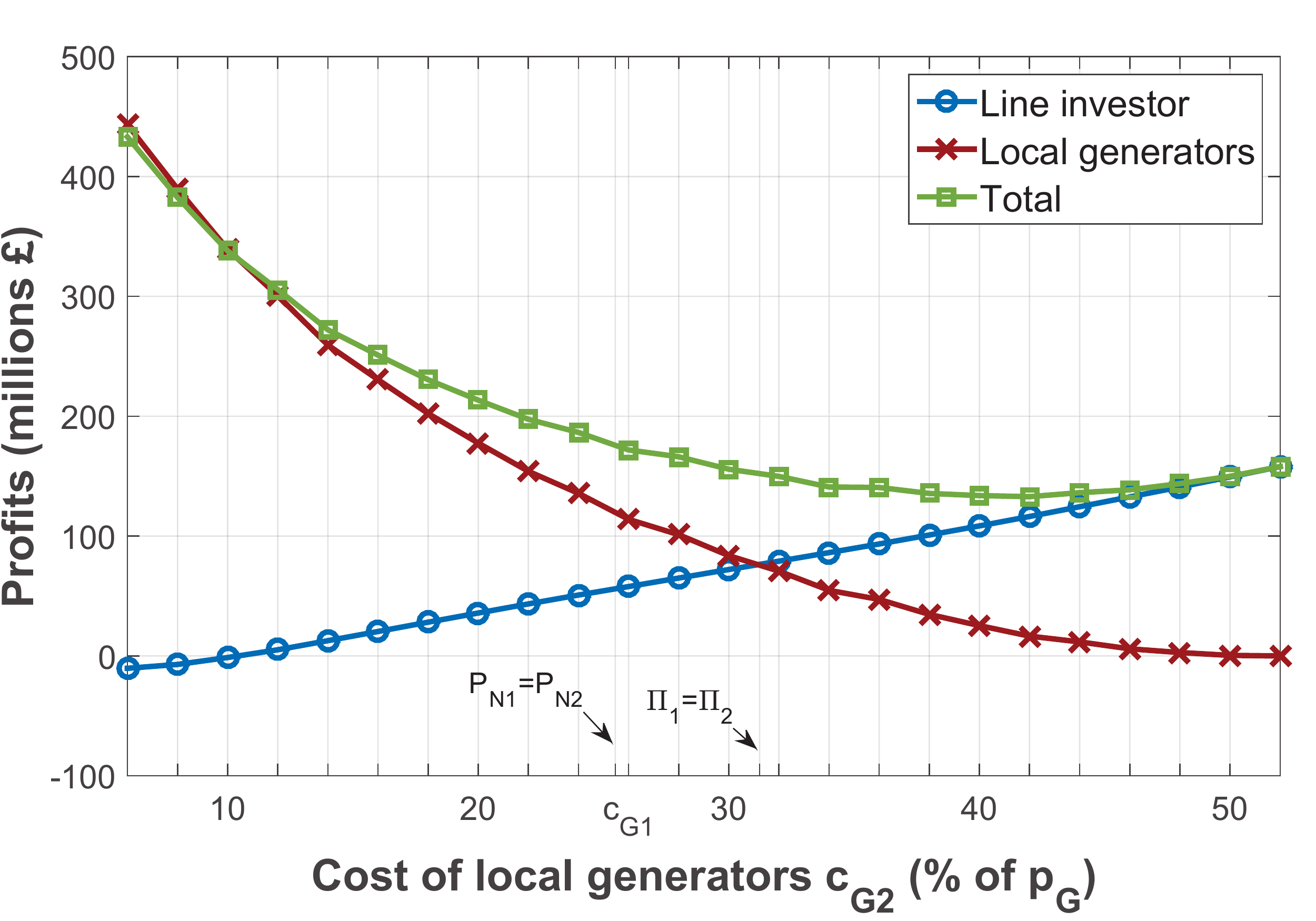}
\end{subfigure}
 & 
 \begin{subfigure}{0.33\textwidth}
 \includegraphics[width=\linewidth]{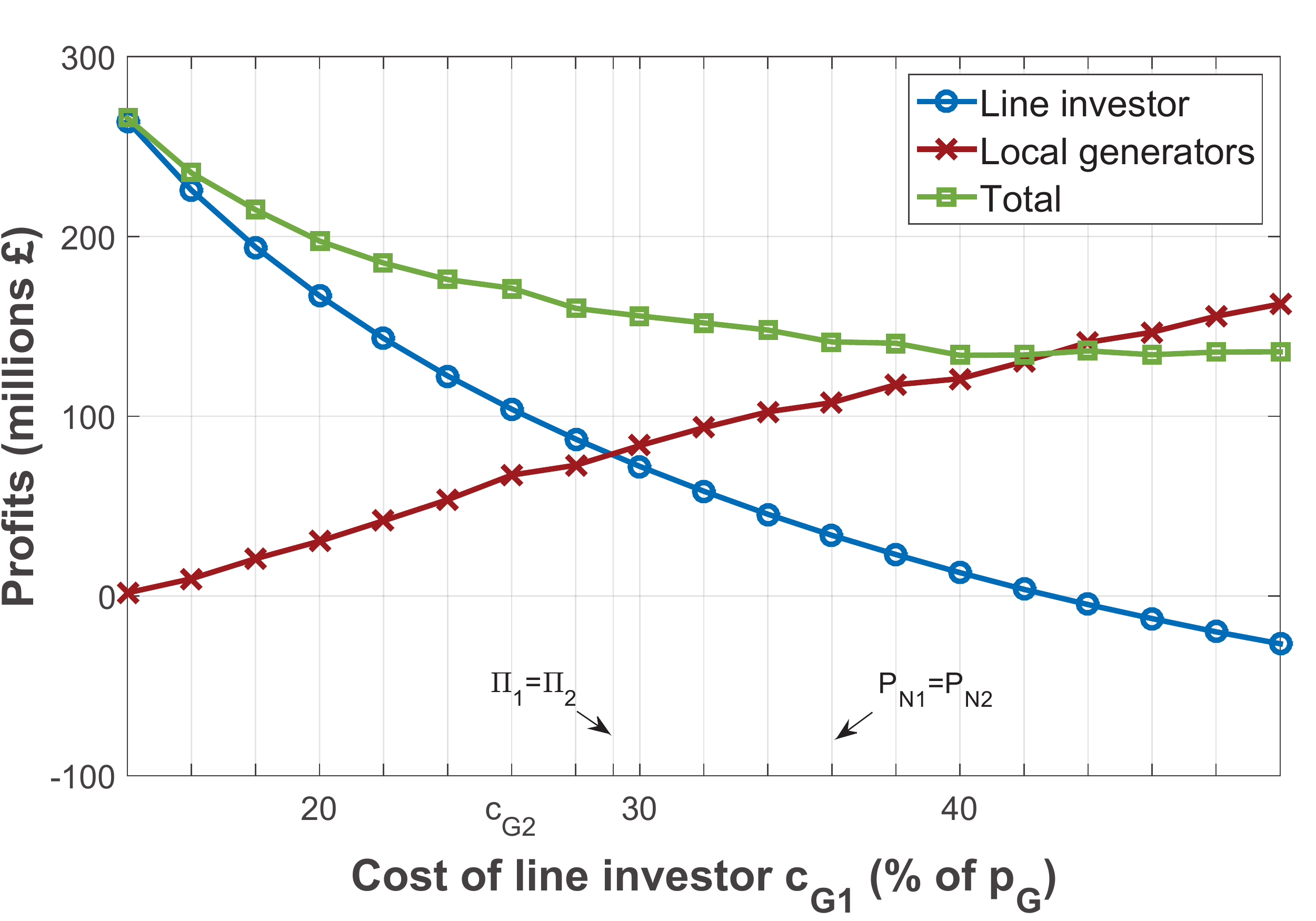}
 \end{subfigure} &
 \begin{subfigure}{0.33\textwidth}
 \includegraphics[width=\linewidth]{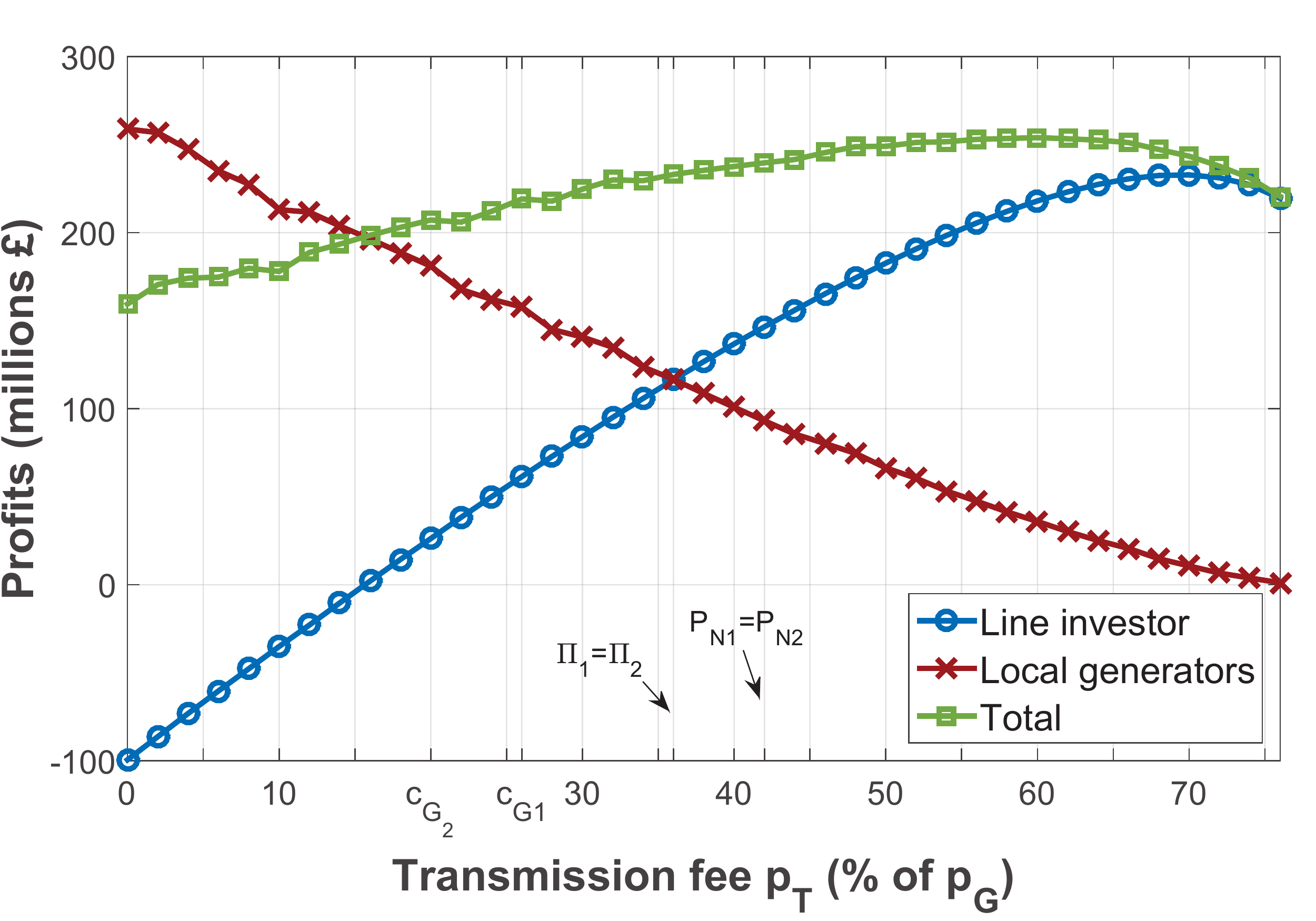}
 \end{subfigure}
 \\
 \small(3) &
 \begin{subfigure}{0.33\textwidth}
\includegraphics[width=\linewidth]{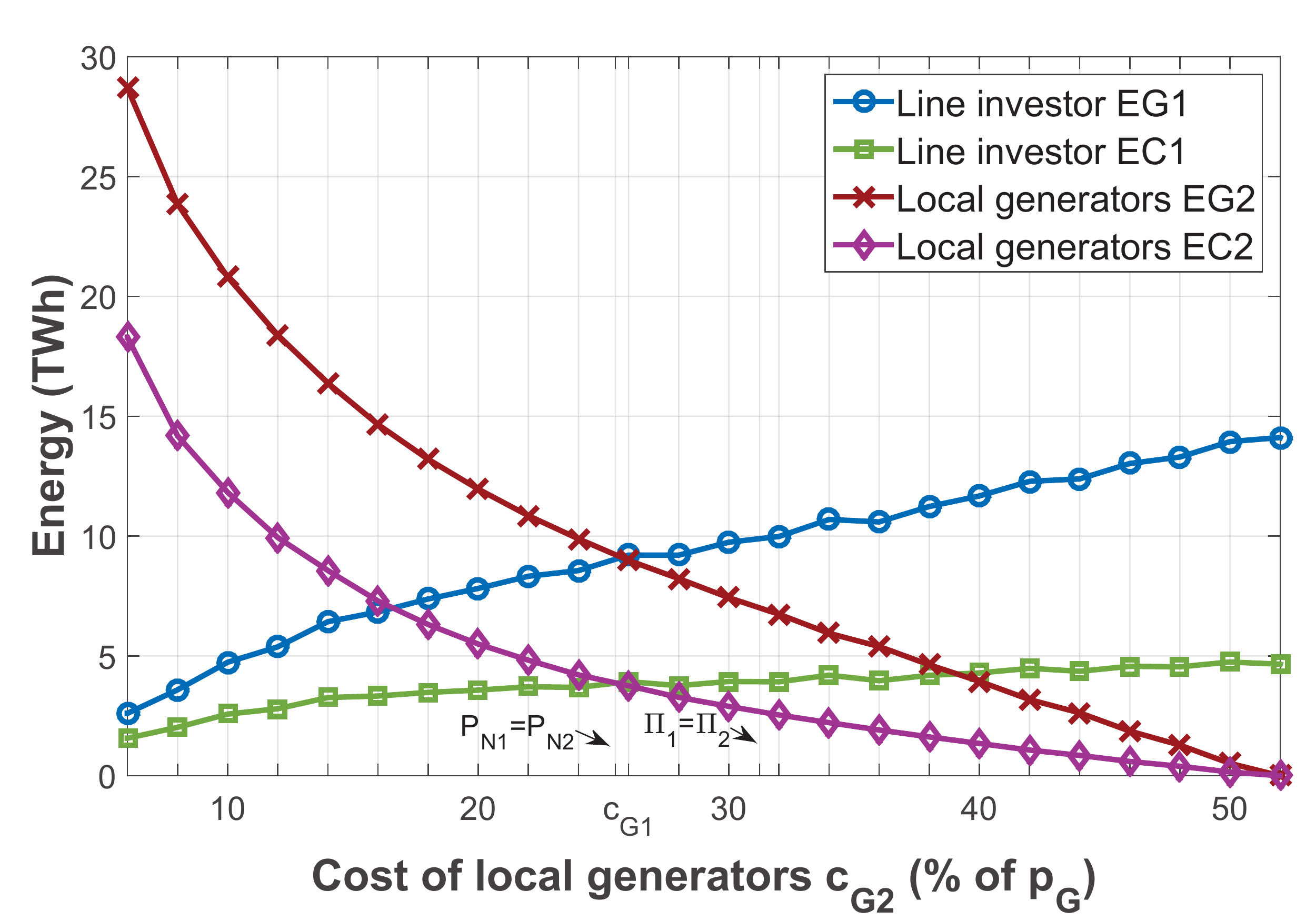} \caption{}
\end{subfigure}
 & 
 \begin{subfigure}{0.33\textwidth}
 \includegraphics[width=\linewidth]{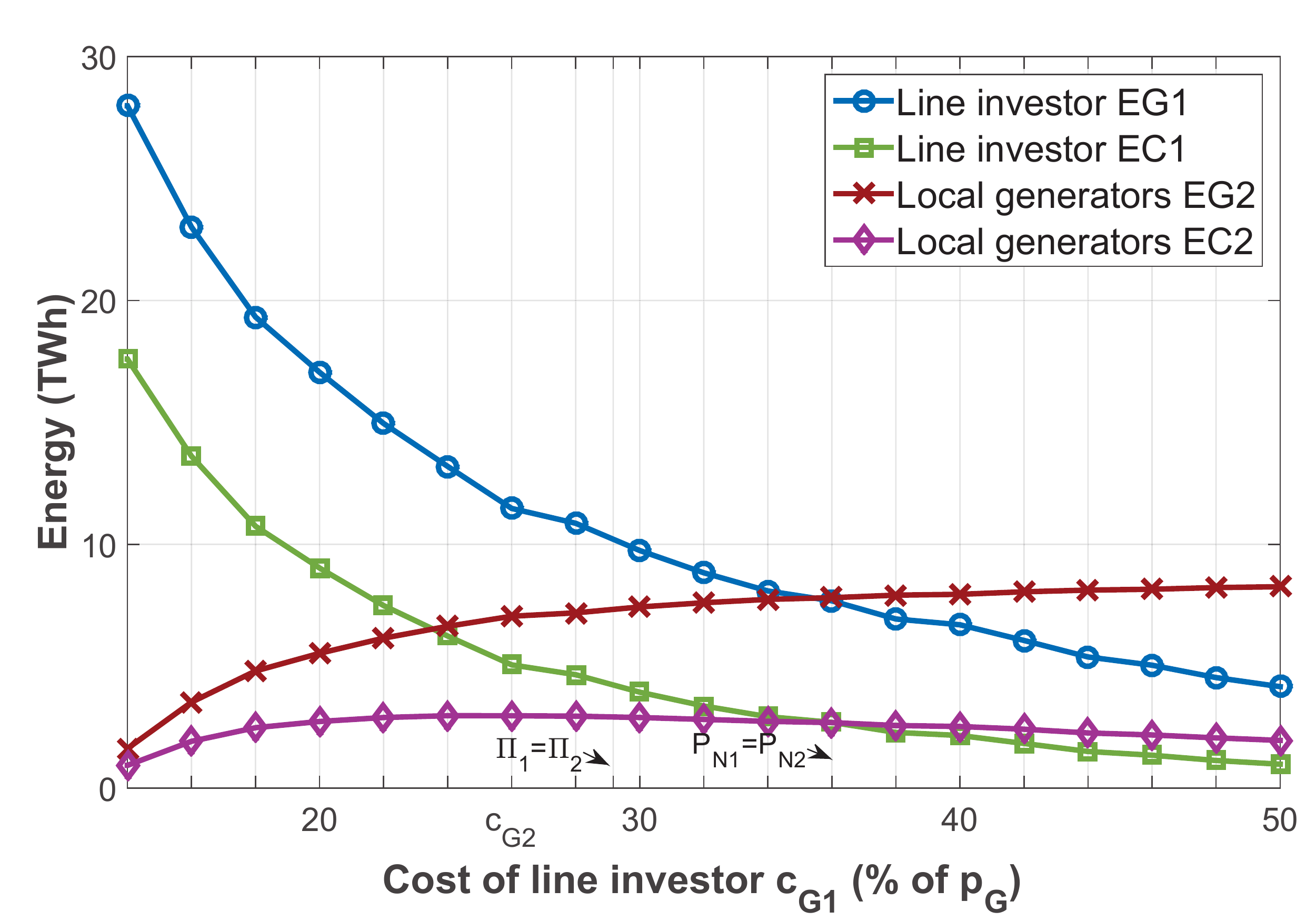} \caption{}
 \end{subfigure} &
 \begin{subfigure}{0.33\textwidth}
 \includegraphics[width=\linewidth]{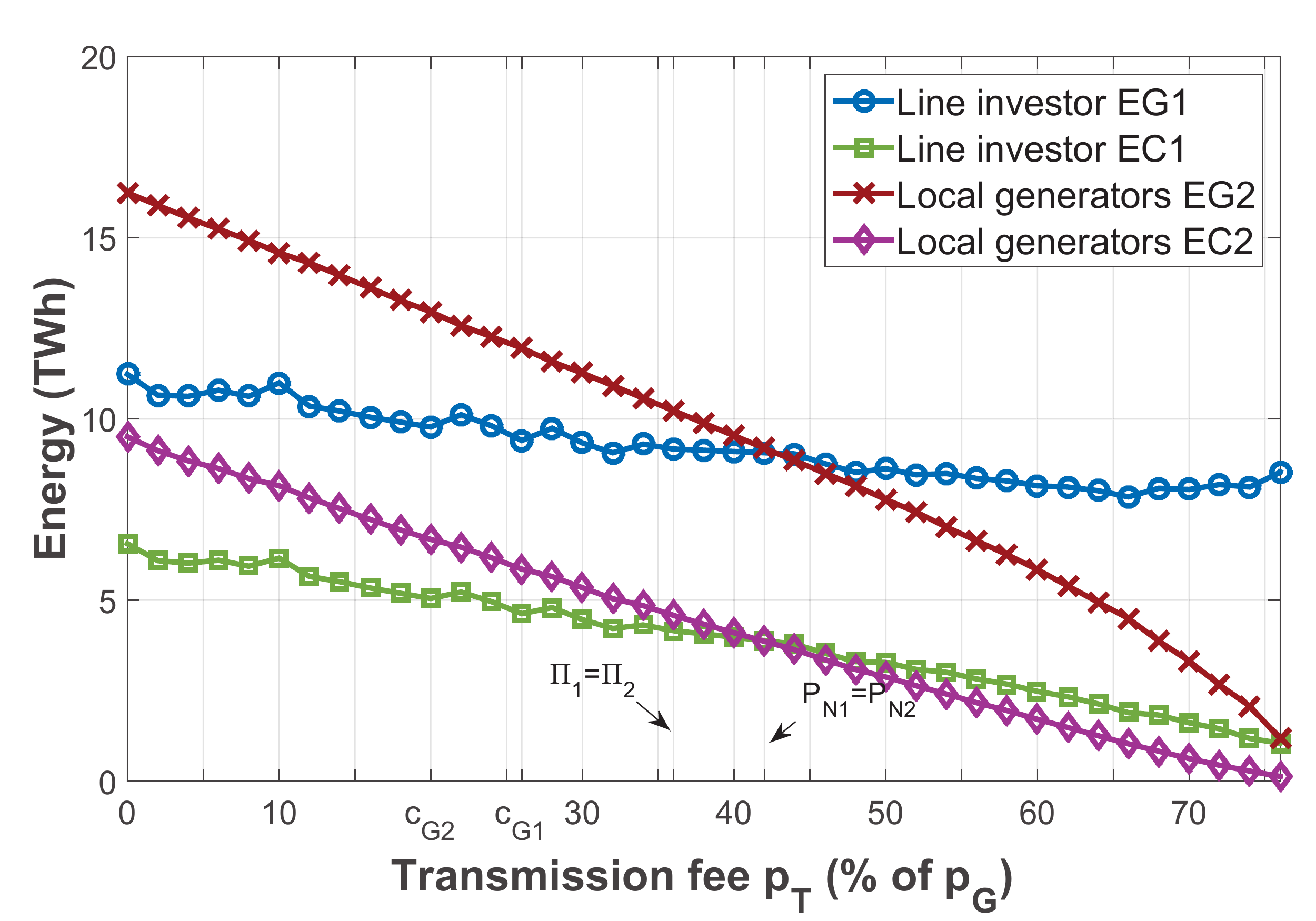} \caption{}
 \end{subfigure} \\
 \hline
\end{tabular}
\caption{Rows (1), (2) and (3) show generation capacity built, profits, energy that could have been generated and energy curtailed at Stackelberg equilibrium, respectively, column (a) shows dependency on local generators' generation cost, (b) on line investor's generation cost and (c) on imposed transmission fee}
\label{Stackel_graphs}
\end{figure*}

Total generation capacity installed decreases as $c_{G_2}$ increases due to the reduction of $P_{N_2}$ installed, as shown in Fig.~\ref{Stackel_graphs}-1a. We can observe two critical points, $c_{G_2}\simeq0.255p_G$ where players install equal generation capacities $P_{N_1}=P_{N_2}$ (Fig.~\ref{Stackel_graphs}-1a) and $c_{G_2}\simeq0.312p_G$, where profits for both players are equal $\Pi_1=\Pi_2$ (Fig.~\ref{Stackel_graphs}-2a). For $c_{G_2}<0.255p_G$, local generators install more generation capacity than the line investor. For $c_{G_2}=0.255p_G$ to $0.312p_G$ although $P_{N_1}>P_{N_2}$ (Fig.~\ref{Stackel_graphs}-1a), the leader's profit is lower, due to the additional cost of installing the line $C_T$. If $c_{G_2}$ increases further, then local generators decrease their installed capacity, which eventually leads to equal profits and sequentially to the leader's  overcoming the follower's profit (Fig.~\ref{Stackel_graphs}-2a).
\end{itemize}

\vspace{-2mm}
\begin{itemize}
\item \textit{Scenario 2: Varying line investor's cost:} In this scenario, we set $c_{G_2}=0.30p_G$ and $p_T=0.26p_G$, while $c_{G_1}=0.14p_G$ to $0.50p_G$. Results are shown in the second column of Fig.~\ref{Stackel_graphs}.

Total generation decreases as $c_{G_1}$ increases with other parameters remaining equal (Fig.~\ref{Stackel_graphs}-1b). For low leader's generation cost, the line investor installs more generation capacity leading to larger profits. However, as $c_{G_1}$ increases less capacity is installed by the line investor. This leads to decreasing profits. At $c_{G_1}\simeq0.292p_G$ the profits of the players become equal (Fig.~\ref{Stackel_graphs}-2b). From $c_{G_1}\simeq0.292p_G$ to $0.36p_G$, the line investor continues to install more capacity up to $c_{G_1}\simeq0.36p_G$, where players install equal generation capacity (Fig.~\ref{Stackel_graphs}-1b).
\end{itemize}
\vspace{-2mm}

\begin{itemize}
\item \textit{Scenario 3: Varying transmission fee:} We assume  that $c_{G_1}=0.26p_G$ and $c_{G_2}=0.20p_G$, while the transmission fee varies from $p_T=0$ to $0.76p_G$. Results are shown in the third column of Fig.~\ref{Stackel_graphs}.

The total generation capacity decreases as $p_T$ increases, due to local generators installing less capacity. The leader's generation capacity is relatively constant with varying $p_T$. Note here that the leader may react to the decreasing capacity of the local generators as $p_T$ increases, both by decreasing or increasing their own built capacity, as $p_T$ increases (Fig.~\ref{Stackel_graphs}-1c).

When the transmission fee is $p_T<0.42p_G$, followers install more capacity as a result of the transmission fee and cheaper generation cost. However, as $p_T$ increases, the revenues drop for local generators, who install less $P_{N_2}$, up to $p_T\simeq0.42p_G$ where players install equal capacities (Fig.~\ref{Stackel_graphs}-1c). Local generators have larger profits until $c_{G_2}\simeq0.36p_G$ where profits break even, mainly due to the high power line installation cost $C_T$ (Fig.~\ref{Stackel_graphs}-2c).

For this setting, the transmission fee needs to be at least $p_T\simeq0.15p_G$. Charging a transmission fee below this amount would make it uneconomical for the line investor to install the line, given the expected response by local generators. Morever, the line investor needs to install roughly as much generation capacity as local generators to achieve similar profit, in this scenario. In contrast, if $p_T$ is set too high, it is not feasible for local generators to invest in renewable energy at this location (Fig.~\ref{Stackel_graphs}-2c).
\end{itemize}
\vspace{-2mm}
\subsection{Discussion of results}

As shown in the results, for every set of cost ($c_{G_1}$, $c_{G_2}$) and revenue parameters ($p_T$, $p_G$), there is an upper limit of total generation capacity being installed at Location B, which is equal to the sum of rated capacities installed by each player. In all sets of scenarios, total capacity decreases as the tested parameter value increases (Fig.~\ref{Stackel_graphs}-1a, Fig.~\ref{Stackel_graphs}-1b and Fig.~\ref{Stackel_graphs}-1c). Each player installs less capacity as their generation cost increases, while the other player benefits by increasing their capacity (Fig.~\ref{Stackel_graphs}-1a and Fig.~\ref{Stackel_graphs}-1b). The cost of local generators has a larger impact on the capacities installed for both players, as shown by comparing Fig.~\ref{Stackel_graphs}-1a and Fig.~\ref{Stackel_graphs}-1b, as local generators face the additional cost of transmission charges. Profits have similar behaviour to the generation capacities built in Scenarios 1 and 2, while in Scenario 3, the line investor's profit increases because of larger revenues from transmission (Fig.~\ref{Stackel_graphs}-2c). Note here that the players profits are not equal when $P_{N_1}=P_{N_2}$ (which over a long time window means  $E_{G_1}\simeq E_{G_2}$ and $E_{C_1}\simeq E_{C_2}$ as shown when comparing Fig.~\ref{Stackel_graphs}-1a with Fig.~\ref{Stackel_graphs}-3a, or Fig.~\ref{Stackel_graphs}-1b with Fig.~\ref{Stackel_graphs}-3b and Fig.~\ref{Stackel_graphs}-1c with Fig.~\ref{Stackel_graphs}-3c), because of transmission charges $p_T$, but also because of different generation costs and $C_T$. 

If the followers' generation cost is much smaller than the line investor's (assuming for example that local generators might have access to cheaper land or favourable licensing approval), then the line investor will need to charge a high transmission fee to have positive earnings (Fig.~\ref{Stackel_graphs}-2c). On the other hand, if the leader's cost is much smaller, the generation capacity will mostly be installed by the line investor, as there is no room for profitable investment of other renewable producers. Moreover, in Scenario 3 it is shown that the followers' generation capacity decreases as $p_T$ increases, but this does not always result in the leader increasing their own capacity (see Fig.~\ref{Stackel_graphs}-1c). Estimating the best response is a complex procedure which depends on the curtailment imposed and varying demand. In a similar way, if we assume that local generators increase their installed capacity, it is possible for the line investor to slightly increase their own generation capacity, as this strategy move may minimise the profit losses incurred, as long as the increased cost of installing the additional generation capacity units leading to larger energy curtailed is counter-balanced by the revenues generated by satisfying a larger demand at times when no curtailment occurs.

As shown in Scenario 3, $p_T\simeq0.15p_G$ is the minimum value of the transmission fee that allows profit for the line investor. Similarly, if the transmission fee is set too high, then local investors will not invest in renewable generation, as their profit diminishes with increasing transmission fee. As $p_T$ is set by the system regulator, this method determines a feasible range that allows both transmission and generation investments to be profitable (Fig.~\ref{Stackel_graphs}-2c).

The model developed in this work can model grid reinforcement projects performed by private investors, who aim to maximise their profits instead of typical cost minimising techniques or maximising social welfare objectives, that exist when network upgrade is performed by the system operators. Typical settings where this model can be applied in practice include numerous locations where demand and generation are not co-located. Finally, the model developed  offers good insights to the strategic game formed between the players, for varying cost parameters. Conclusions can be reached either directly on real data measurements or their distributions.

\section{Conclusions \& Future work}
\label{Con_Fut}

In this work we have shown how privately developed network upgrade for RES connection to the electricity grid can lead to a leader-follower game between the line investor and local investors. Curtailment and line access rules play a key role in the strategic game, the equilibrium of which can be used to determine optimal generation capacities installed in such settings and their associated profits. The model developed can capture the stochastic nature of renewables and the variation in demand. We have identified the equilibrium of the game and have shown how the optimal solution depends on the generation costs of the two players and the transmission fee. Most crucially, the latter can be used by regulators to calculate a feasible range for the transmission fee, that allows both network upgrade and local renewable generation investment. We have developed a methodology for the equilibrium of the game that utilises real data, both on the supply and demand side and have applied this to a case study in Western Scotland. We used a big dataset analysis that spans over the course of 17 years. Other contributions of our work include a study on different curtailment rules and their effects on the capacity factor of wind generators, hence their profitability and viability of investment. Finally, we have proposed a new curtailment rule which ensures equal share of curtailment amongst generators with minimal disruption. In the future, we plan to extend the model to multi-location settings and dispatch decisions that include flexibility on the demand side, such as demand response or energy storage devices, which can be used to partially defer curtailment. A combined system with partial storage, demand response and rare curtailment events could be the   most realistic solution in practice.

\section*{Acknowledgements}
The authors would like to thank Community Energy Scotland, SSE, the National Grid and the UK Meteorological Office for data access and all the information provided. Valentin Robu and David Flynn would like to acknowledge the support of the EPSRC National Centre for Energy Systems Integration (CESI) [EP/P001173/1].

\section*{References}
\bibliographystyle{elsarticle-num} \bibliography{mybiblio}

\end{document}